\documentclass[preprintnumbers,article,amsmath,amssymb,floatfix,10pt,prd,onecolumn,
superscriptaddress,nofootinbib]{revtex4-2}
\usepackage{bm}
\usepackage{amsfonts}
\usepackage{latexsym}
\usepackage[latin1]{inputenc}
\usepackage{graphicx}
\usepackage{amsmath}
\usepackage{palatino}
\usepackage{mathpazo}
\usepackage{textcomp}
\usepackage{float}
\usepackage{booktabs}
\usepackage{dcolumn}
\usepackage{ragged2e}
\usepackage{hyperref}
\hypersetup{colorlinks,citecolor=blue}
\hypersetup{colorlinks=true,linkcolor=red,filecolor=magenta,    urlcolor=blue}
\usepackage{amsmath}
\usepackage{xcolor}
\usepackage{color}
\usepackage{orcidlink}
\usepackage{epsfig}
\usepackage{caption}
\usepackage{subcaption}
\usepackage{commath}
\def\jnl@style{\it}
\def\aaref@jnl#1{{\jnl@style#1}}

\def\aaref@jnl#1{{\jnl@style#1}}

\def\aj{\aaref@jnl{AJ}}                   
\def\apj{\aaref@jnl{ApJ}}                 
\def\apjl{\aaref@jnl{ApJ}}                
\def\apjs{\aaref@jnl{ApJS}}               
\def\apss{\aaref@jnl{Ap\&SS}}             
\def\aap{\aaref@jnl{A\&A}}                
\def\aapr{\aaref@jnl{A\&A~Rev.}}          
\def\aaps{\aaref@jnl{A\&AS}}              
\def\mnras{\aaref@jnl{Mon.~Not.~Roy.~Astron.~Soc.}}             
\def\prd{\aaref@jnl{Phys.~Rev.~D}}        
\def\prc{\aaref@jnl{Phys.~Rev.~C}}  
\def\prl{\aaref@jnl{Phys.~Rev.~Lett.}}    
\def\qjras{\aaref@jnl{QJRAS}}             
\def\skytel{\aaref@jnl{S\&T}}             
\def\ssr{\aaref@jnl{Space~Sci.~Rev.}}     
\def\zap{\aaref@jnl{ZAp}}                 
\def\nat{\aaref@jnl{Nature}}              
\def\aplett{\aaref@jnl{Astrophys.~Lett.}} 
\def\apspr{\aaref@jnl{Astrophys.~Space~Phys.~Res.}} 
\def\physrep{\aaref@jnl{Phys.~Rep.}}      
\def\physscr{\aaref@jnl{Phys.~Scr}}       
\def\commat{\aaref@jnl{Comm.~Math.~Phys.}}              
\def\science{\aaref@jnl{Science}}               
\def\cqg{\aaref@jnl{Classical Quant.~Grav.}}            
\def\jpcs{\aaref@jnl{JPCS}}                                     
\def\ijmpd{\aaref@jnl{Int.~J.~Mod.~Phys.~D}}                    
\def\grg{\aaref@jnl{Gen.~Relat.~Gravit.}}               
\def\rpp{\aaref@jnl{Rep.~Prog.~Phys.}}          
\def\npa{\aaref@jnl{Nucl.~Phys.~A}}        
\def\lrr{\aaref@jnl{Living Rev.~Rel.}}                   
\def\jcap{\aaref@jnl{J.~Cosmology Astropart.~Phys.}}    
\def\rmp{\aaref@jnl{Rev.~Mod.~Phys.}}   
\def\epjc{\aaref@jnl{Eur.~Phys.~J.~C}} 
\def\plb{\aaref@jnl{~Phy.~Lett.~B}} 
\def\mpla{\aaref@jnl{Mod.~Phy.~Lett.~A}} 
\def\arxiv{\aaref@jnl{arxiv.org}}

\allowdisplaybreaks[1]

\begin{document}
\color{black}  

\title{Modeling and Analyzing Stability of Hybrid Stars within $f(Q)$ Gravity}

\author{Piyali Bhar}
\email{piyalibhar90@gmail.com}
\affiliation {Department of Mathematics, Government General Degree College Singur, Hooghly,
 West Bengal 712409, India}

\author{M.R. Shahzad}
\email{rezwan.chaudhery@gmail.com}
\affiliation {Department of Mathematics, Bahauddin Zakariya University Multan, Vehari Campus, Vehari 61100, Pakistan}

 \author{Sanjay Mandal}
\email{sanjaymandal960@gmail.com}
\affiliation{Faculty of Symbiotic Systems Science, Fukushima University, Fukushima 960-1296, Japan}

\author{P.K. Sahoo}
\email{pksahoo@hyderabad.bits-pilani.ac.in}
\affiliation {Department of Mathematics, Birla Institute of Technology and Science-Pilani, Hyderabad Campus, Hyderabad 500078, India}

\begin{abstract}
This study uses the Krori-Barua type metric to represent hybrid stars within the $f(Q)$ theory of gravity. We postulate that the hybrid star also contains strange quark matter in addition to regular baryonic matter. To investigate the physical viability of the hybrid star model, we present the graphical behavior of the energy density, radial pressure, and tangential pressure, equation of state parameters, anisotropy, and stability analysis, respectively, by choosing the compact star EXO 1785-248 with a mass of $1.3_{-0.2}^{+0.2}~M_{\odot}$ and radius  $8.849_{-0.4}^{+0.4}$ km with five different values of the coupling constants as $a=2$, $a=4$, $a=6$, $a=8$, and $a=10$. The maximum allowed masses and corresponding radii have been calculated using the $M-R$ curve for three different coupling parameter '$a$' values which match the observational data of three distinct compact stars, namely LMC X-4, SMC X-1, and 4U 1538-52. So, the $f(Q)$ theory of gravity can provide results that are plausible for describing the macroscopical characteristics of hybrid star candidates.
\end{abstract}

\maketitle

\section{Introduction}
General relativity (GR) has been and continues to be a fruitful and useful theory to explain gravitational interaction.
However, the evidence from observations exposing the rapid expansion of the Universe put the main theoretical issues on GR and provided abundant information concerning the evolutionary phase of the Universe and its hidden secrets \cite{r1}-\cite{r15}.
The accelerated expansion of the cosmos is a crucial factor in the dynamic history of the universe. The idea of what unknown power is behind this acceleration is intriguing and open-ended. There is currently no competing hypothesis within the framework of GR that can adequately explain this mystery, prompting scientists to seek out new explanations. 

Theoretically, there are two ways to explain the present phase of the accelerating expansion of the universe. The first strategy is to alter the universe's composition by introducing new matter and energy components, such as DE, which is currently concealed due to high negative pressure. Einstein incorporated the well-known cosmological constant ($\Lambda$) into GR field equations, and it is the most favored option for DE, given that it is consistent with observation. According to the hypothesis, the cosmological constant arose from the vacuum energy anticipated by quantum field theory \cite{r16}. However, this well-motivated candidate for DE, the cosmological constant ($\Lambda$), has some flaws. The main one is its constant equation of state. Before the discovery of late-time accelerated expansion, numerous dynamical models of $\Lambda$ were investigated in the literature in an attempt to address the cosmological constant problem. The cosmological constant in the Einstein field equations (EFEs) was introduced after the breakthrough of cosmic acceleration. In addition, the case of a cosmological constant can be reduced to that of a slow-roll scalar field (a potential-dominated scalar field). Due to its dynamical equation of state, the quintessence model \cite{r17} can be easily applied when considering a scalar field.  K-essence \cite{r18}, Phantom DE \cite{r19}, Chameleon \cite{r20}, Chaplygin gas \cite{r21,r22}, and Tachyon \cite{r23} are only a few of the dynamical (time-varying) models of DE that have been propounded. Along with the DE scenario of the universe, black hole physics seeks a lot of researchers to into it in modern astrophysics (for instance, one can see some of the interesting studies on black holes \cite{oa1,oa2,oa3,oa4}).

The alternative avenue for elucidating the present acceleration of the universe's expansion involves the adjustment of the Einstein-Hilbert action within the framework of GR. To this end, Qadir et al. \cite{r23a} looked into the many characteristics of modified relativistic dynamics and their impact, arguing that the standard GR may need to be revised in order to resolve a number of cosmological challenges, including the dark matter and quantum gravity problems. Additionally, the LIGO detector's discovery of gravitational waves is seen as a useful tool for distinguishing between various extended theories of gravity. The foundational principle within the framework of GR resides in the incorporation of curvature derived from Riemannian geometry, which is mathematically encapsulated by the Ricci scalar $R$. The modified $f(R)$ gravity theory can be regarded as a straightforward alteration of the GR framework, wherein the Ricci scalar R is substituted with a more comprehensive function of R \cite{r24}. Moreover, it is worth noting the existence of alternative frameworks to GR, such as the teleparallel equivalent of GR (TEGR). Within TEGR, the intricate dynamics of gravitational interactions are elegantly captured through the incorporation of torsion T as a fundamental concept. In the framework of GR, the Levi-Civita connection is intimately linked to the concept of curvature, yet it exhibits a remarkable absence of torsion. Conversely, within the realm of teleparallelism, the Weitzenbock connection is intricately tied to torsion while maintaining a notable absence of curvature \cite{r25,r25a}. Similarly, as in the case of $f(R)$ gravity, the $f(T)$ gravity theory can be regarded as the most elementary alteration of the Teleparallel Equivalent of General Relativity (TEGR). 

Recently, a new equivalent to GR theory has been proposed called symmetric teleparallel equivalent to GR, usually abbreviated as STGR, in which gravity is incorporated by the non-metricity $Q$ while taking curvature and torsion equal to zero \cite{r26,r27,r28,snehafqt}. Subsequently, the modification of STGR has been introduced similarly to the $f(R)$ and $f(T)$ discussed above. It is called the $f(Q)$ theory of gravity \cite{r26}, also called symmetric teleparallel gravity. This recently developed theory drastically attracted numerous researchers to unveil the intriguing consequences in cosmological and astrophysical scenarios. 

Lazkoz et al. \cite{r30} have made a noteworthy contribution to the field of symmetric teleparallel gravity, wherein they have imposed constraints on a collection of $f(Q)$ functions. To accomplish this, the Lagrangian $f(Q)$ was redefined in terms of the redshift z, and a comprehensive analysis was conducted to examine the empirical limitations. Solanki et al. \cite{r30a} show that the $f(Q)$ gravity model adequately addresses the dark energy problem which can not be dealt with by the ordinary GR. Mandal et al. \cite{r31} conducted a pertinent investigation on $f(Q)$ gravity, aiming to comprehend its behavior through the lens of energy conditions. The study involved deriving energy conditions specific to $f(Q)$ gravity and examining the self-stability of two $f(Q)$ models. Furthermore, the researchers employed the parametrization technique alongside the cosmographic concept to confine three sets of cosmographic functions using statistical analysis in the realm of $f(Q)$ gravity, utilizing the extensive Pantheon supernovae data \cite{r32}. Indeed, a plethora of intriguing investigations have been conducted employing $f(Q)$ theory (refer to the comprehensive elucidation in references \cite{r27,r33,r33a,r34,r36,r37}.

Zhao \cite{r38} highlighted that on certain occasions, the selection of a gauge might clash with the coordinate system we have chosen, predicated upon the principles of symmetry. To surmount this quandary, the author devised the $f(Q)$ theory in a covariant manner, thereby enabling us to ascertain an appropriate non-vanishing affine connection for a prescribed metric. Further, the author employed this methodology to examine two significant scenarios: the isotropic homogeneous expanding universe and the spherically symmetric static spacetime. In their seminal work, Hu et al. \cite{r39} elegantly showcased the Hamiltonian analysis of $f(Q)$ gravity while imposing the coincident gauge condition as a constraint. By employing the conventional Dirac-Bergmann algorithm, it has been demonstrated that within the framework of $f(Q)$ gravity, a total of 8 distinct physical degrees of freedom emerge. The observed outcome indicated that the diffeomorphism symmetry of $f(Q)$ gravity is entirely compromised as a consequence of the gauge fixing. Gadbail et al. \cite{r40} introduced a captivating array of explicit reconstructions of the $f(Q)$ gravity, derived from the backdrop of FLRW evolution history. The discovery pertains to the identification of the broader functional properties of the non-metricity scalar $Q$, which exhibit precise adherence to the exact accelerated cosmic expansion history, as described by the $\Lambda$CDM model. De and Loo \cite{r41} examined the viability of the $f(Q)$ models and argued that certain non-linear models violate the energy constraints. Mandal and his collaborator studied the limitation of the EOS parameter using different observational restrictions \cite{r41a} and the influence of the bulk viscosity on the cosmological models \cite{r41b}. The cosmological scenario and $f(Q)$ as a suitable alternative to the $\Lambda$CDM model has been studied in \cite{r41c}.
Recently, Koussour et al. \cite{r42} introduced a novel parametrization for the Hubble parameter, employing a model-independent approach. Subsequently, they adeptly applied this parametrization to the Friedmann equations within the framework of the FLRW Universe. The optimal values of the model parameters are approximated through the utilization of the amalgamated datasets encompassing the updated Hubble parameter measurements $(H(z))$, the Pantheon, and the Baryon Acoustic Oscillation (BAO) datasets corresponding to different data points. This estimation is achieved by employing the Markov Chain Monte Carlo (MCMC) technique. Ultimately, it has been determined that the theoretical framework substantiates the notion of the currently observed expansion of the universe, with the equation of state parameters exhibiting characteristics akin to those of the quintessence model. In the realm of astrophysics, it is customary to regard compact entities, such as neutron stars, quark stars, and black holes, as experimental arenas wherein one may scrutinize specific attributes pertaining to the gravitational fields. Relativistic stellar objects have garnered significant attention from researchers over the past few decades. In the modified $f(Q)$ gravity, the stellar structures like compacts stars and wormhole geometries have been studied in \cite{r43,r44,r45}

Motivated by the above consequences of the newly established $f(Q)$ theory, we studied the dark energy star models in this novel theory. The sequence of our study is as follows: The basic field equations and a short discussion on $f(Q)$ gravity have been covered in section~\ref{sec2}. The internal spacetime is addressed in Section~\ref{sec3}. The fundamental field equations have been solved using the KB metric in the next section. In section~\ref{sec5}, it has been assessed that the inner spacetime matches the exterior Schwarzschild vacuum solution at the boundary $r = R$ and the constants of metric coefficients are obtained in terms of mass and radius of the star. $M-R$ diagrams have been used to determine the maximum permissible masses and corresponding radii for three distinct values of the coupling parameter `a' in section~\ref{sec6}. The next section discusses some of the model's physical characteristics, and section~\ref{sec8} provides some concluding notes.
\section{$f(Q)$ gravity and basic field equations}\label{sec2}

We consider the action for $f(Q)$ gravity given by \cite{r29},
\begin{eqnarray}\label{r1}
S=\int\left[\frac{1}{2}f(Q)+\mathcal{L}_m\right]\sqrt{-g}d^{4}x,
\end{eqnarray}
In this modification, we encounter a scenario where $f(Q)$ embodies a comprehensive function of $Q$, while $g$ symbolizes the determinant of the metric $g_{\mu \nu}$. Additionally, we have the matter Lagrangian density denoted as $\mathcal{L}_m$.

The non-metricity tensor is defined to be
\begin{eqnarray}\label{r2}
Q_{\alpha\beta\gamma}=\nabla_\alpha g_{\beta\gamma}=-L^{\lambda}_{\alpha\beta}g_{\lambda\gamma}-L^{\lambda}_{\alpha\gamma}g_{\lambda\beta},
\end{eqnarray}
 wherein the deformation term regarded as,
\begin{eqnarray}\label{r3}
L^{\alpha}_{\beta\gamma}=\frac{Q^\alpha_{\beta\gamma}}{2}-Q_{(\beta\gamma)}^{    \alpha},
\end{eqnarray}
and the non-metricity tensor can be reduced to two independent traces as
\begin{eqnarray}\label{r4}
Q_\alpha={Q_\alpha}^\beta ~_\beta, ~~~~ \tilde{Q}_\alpha= {Q^\beta}_{\alpha\beta}.
\end{eqnarray}

The super-potential tensor in terms of the non-metricity tensor obtained as

\begin{eqnarray}\label{r5}
P^\alpha_{\beta\gamma}=-{Q^\alpha}_{\beta\gamma}+2Q^\alpha_{(\beta \gamma)}-Q^\alpha g_{\beta\gamma}-\tilde{Q}^\alpha g_{\beta\gamma}-\delta^\alpha_{({\beta ^{Q_\gamma}})},
\end{eqnarray}

the non-metricity scalar can be obtained by taking a trace of the non-metricity tensor $Q_{\alpha\beta\gamma}$ as \cite{r29}

\begin{eqnarray}\label{r6}
Q=-Q_{\alpha\beta\gamma} P^{\alpha\beta\gamma}.
\end{eqnarray}

Now, by variation of the action (\ref{r1}) with respect to the metric tensor $g_{\alpha\beta}$ the field equations can be obtained as

\begin{eqnarray}\label{r7}
\frac{2}{\sqrt{-g}}\nabla_\alpha (\sqrt{-g} f_Q P^\alpha_{\beta\gamma})+\frac{1}{2} g_{\beta\gamma} f +f_Q(P_{\beta\alpha\lambda} Q^{\alpha\lambda}_\gamma-2Q_{\alpha\lambda\beta} P^{\alpha\lambda}_\gamma)=-T_{\beta\gamma},
\end{eqnarray}
where $f_Q$ represents the derivative of $f$ with respect to $Q$. Moreover, by varying (\ref{r1}) concerning the affine connection, one can obtain

\begin{eqnarray}\label{r8}
\nabla_\beta \nabla_\gamma (\sqrt{-g} f_Q P^{\beta\gamma}_{~~~\lambda})=0.
\end{eqnarray}

\section{Interior Spacetime}\label{sec3}
We make use of the space-time of a static, self-gravitating sphere supplied by the line element
\begin{equation}\label{line1}
ds^{2}_-=e^{\nu}dt^{2}-e^{\lambda}dr^{2}-r^{2}(d\theta^{2}+\sin^{2}\theta d\phi^{2}),
\end{equation}
to model a star. Where $\nu$ and $\lambda$ are metric potentials that only depend on the radial coordinate `$r$'. In such a manner, the physical interior of the self-gravitating object can be represented in the co-moving frame as a relativistic anisotropic fluid with density $\rho$, radial pressure $p_r$, and tangential pressure $p_t$.
In this study, a hybrid star model containing strange quarks and normal baryonic matter is presented. The hybrid star was motivated by the idea of the transition of phase from quark matter to nuclear matter \cite{Yan:2012mk, Schertler:1997za, Schertler:2000xq}.
The two-fluid model's associated energy-momentum tensor is represented by the following notation:
  \begin{eqnarray}
  T_0^0=\rho+\rho_q ,\label{t1}\\
  T_1^1=-(p_r+p_q),\\
  \text{and}~~~
  T_2^2=T_3^3=-(p_t+p_q).\label{t3}
\end{eqnarray}
Here, $\rho_q$ and $p_q$ respectively, denote the matter density and pressure related to the strange quark matter.\\
We have the following field equations for a hybrid star in $f(Q)$ gravity using all the aforementioned expressions:
\begin{eqnarray}
\kappa(\rho+\rho_q)&=&\frac{e^{-\lambda}}{2r^2}\Big[2rf_{QQ}Q'(e^{\lambda} - 1) +
   f_Q\Big((e^{\lambda} - 1)(2 + r\nu') + (1 + e^{\lambda})r \lambda'\Big) +
   f r^2 e^{\lambda}\Big],   \label{fe1}\\
\kappa (p_r+p_q) &=&-\frac{e^{-\lambda}}{2r^2}\Big[2rf_{QQ}Q'(e^{\lambda} - 1) +
   f_Q\Big((e^{\lambda} - 1)(2 + r \lambda' + r\nu') - 2r\nu'\Big) +
   fr^2e^{\lambda}\Big], \label{fe2}\\
\kappa (p_t+p_q) &=& -\frac{e^{-\lambda}}{4r}\Big[-2rf_{QQ}Q'\nu' +
   f_Q\Big(2\nu'(e^{\lambda} - 2) - r\nu'^2 + \lambda'(2e^{\lambda} + r\nu') -
      2r\nu''\Big) + 2fre^{\lambda}\Big].\label{fe3}
\end{eqnarray}
where $\kappa=8\pi$ and derivative with respect to the radial co-ordinate `$r$' is represented by $()^{\prime}$.\par
Let us use a quadratic form for $f(Q)$ gravity to describe the hybrid star. This function is written as follows:
\begin{eqnarray}\label{g3}
f(Q)=Q+aQ^2,
\end{eqnarray}
where `$a$' is the coupling constant of $f(Q)$ gravity. \\
$Q$ is the non-metricity scalar defined as,
\begin{eqnarray}\label{g2}
Q=\frac{1}{r}(\nu'+\lambda')(e^{-\lambda}-1).
\end{eqnarray}
In the next section, we shall solve the field equations to obtain the model of a hybrid star in $f(Q)$ gravity.

\section{Model of Hybrid Star}\label{sec4}
Let us assume the metric coefficients suggested by Krori-Barua \cite{Krori1975} as follows for our current article to solve the field equations given in (\ref{fe1})-(\ref{fe3}):
\begin{eqnarray}\label{g1}
\lambda = Ar^2,\, \nu = Br^2 + C.
\end{eqnarray}
The above {\em ansatz} has three unknown constants, $A,\, B$ and $C$. With the help of the boundary condition, the expression of these three unknowns will be obtained in the next section.\par
It is believed that when a star runs out of fuel, it eventually forms a compact object that can take the shape of a black hole, neutron star, or white dwarf, depending on how massive it is. The densest and most compact stars are neutron stars. They usually compress a solar mass into a compact radius of roughly 10 km. with densities up to several times that of nuclear matter. 
With such densities in the core, they themselves can take on other forms; for instance, they could be made of condensed mesons or normal nuclear matter in addition to hyperons. At such densities, matter may go through a phase transition, causing nucleons to split into quarks and the formation of deconfined strange quark matter (SQM). If this is the case, compact stars could have SQM in their cores, which would be surrounded by normal baryonic matter and form hybrid stars.
\\
Let us suppose that, for normal baryonic matter, the radial pressure $p_r$ and the matter density $\rho$ are connected by a linear equation of state to solve the field equations (\ref{fe1})-(\ref{fe3}) as
 i.e.,\begin{eqnarray}\label{eos1}
                                                     p_r &=& \alpha \rho-\beta,
                                                   \end{eqnarray}
                                                   the constants $\alpha$ and $\beta$ both are positive and $\alpha ~\in$ [0,1] with $\alpha\neq 1/3$. Let us further assume that the MIT bag model equation of state \cite{Cheng1998, Witten1984} provides the pressure-matter density connection for quark matter given as
\begin{eqnarray}\label{eos2}
  p_q &=& \frac{1}{3}(\rho_q-4B_g),
\end{eqnarray}
where $B_g$ is the bag constant of units MeV/fm$^3$ \cite{chodos}. Witten \cite{Witten1984} established the theoretical notion that the strange quark matter could be the true ground state of strongly interacting matter. If this hypothesis is true, compact stars would contain strange matter. It was demonstrated by Farhi and Jaffe \cite{Farhi:1984qu} that the Witten conjecture is true for massless and non-interacting quarks for a bag constant generally ranging from $57$ to $94$ MeV/fm$^3$. Motivated by this, we take into consideration $B_g = 65 MeV/fm^3$ for our current investigation of hybrid stars.
We are now in a position to solve the field equations and obtain the expressions for matter density and pressures for normal baryonic matter as follows:
\begin{multline}
    \rho= \frac{e^{-2 A r^2}}{ \kappa  r^2 (3 \alpha -1)} \large[ (-2 a A^2 r^4-12 a A^2 r^3-4 a A B r^4+8 a A B r^3+16 a A r-2 a B^2 r^4+20 a B^2 r^3+16 a B r)\\+ 2e^{ A r^2}(2 a A^2 r^4+2 a A^2 r^3+4 a A B r^4-12 a A B r^3-16 a A r+2 a B^2 r^4-14 a B^2 r^3-16 a B r-A r^3+A r^2-B r^3+5 B r^2+2)\\+ e^{2 A r^2}(-2 a A^2 r^4+8 a A^2 r^3-4 a A B r^4+16 a A B r^3+16 a A r-2 a B^2 r^4+8 a B^2 r^3+16 a B r\\+2 A r^3-4 A r^2+2 B r^3-4 B r^2+4 B_g \kappa  r^2+3 \beta  \kappa  r^2-4) \large],
\end{multline}

\begin{multline}
    p_r= \frac{e^{-2 A r^2}}{(3 \alpha -1) \kappa  r^2}[e^{2 A r^2} (-2 a \alpha  A^2 r^4+8 a \alpha  A^2 r^3-4 a \alpha  A B r^4+16 a \alpha  A B r^3+16 a \alpha  A r-2 a \alpha  B^2 r^4\\
    +8 a \alpha  B^2 r^3+16 a \alpha  B r-4 \alpha +2 \alpha  A r^3-4 \alpha  A r^2+2 \alpha  B r^3-4 \alpha  B r^2+4 \alpha  B_g \kappa  r^2+\beta  \kappa  r^2)\\
    +2 \alpha  e^{A r^2} \left(2 a A^2 r^4+2 a A^2 r^3+4 a A B r^4-12 a A B r^3-16 a A r+2 a B^2 r^4-14 a B^2 r^3-16 a B r-A r^3+A r^2-B r^3+5 B r^2+2\right)\\
    -2 a \alpha  r (A+B) \left(A r^3+6 A r^2+B r^3-10 B r^2-8\right)],
\end{multline}
\begin{multline}
    p_t=\frac{e^{-2 A r^2}}{(3 \alpha -1) \kappa  r^2}[e^{2 A r^2} (-2 a \alpha  A^2 r^4+8 a \alpha  A^2 r^3-4 a \alpha  A B r^4+16 a \alpha  A B r^3+4 a \alpha  A r+4 a A r-2 a \alpha  B^2 r^4+8 a \alpha  B^2 r^3\\
    +4 a \alpha  B r+4 a B r-\alpha +2 \alpha  A r^3-4 \alpha  A r^2+2 \alpha  B r^3-4 \alpha  B r^2+4 \alpha  B_g \kappa  r^2+\beta  \kappa  r^2-1)+e^{A r^2} (6 a \alpha  A^2 B r^5-2 a A^2 B r^5\\
    +4 a \alpha  A^2 r^4-2 a \alpha  A^2 r^3+2 a A^2 r^3+8 a \alpha  A B r^4-36 a \alpha  A B r^3+4 a A B r^3-8 a \alpha  A r-8 a A r-6 a \alpha  B^3 r^5+4 a \alpha  B^2 r^4-34 a \alpha  B^2 r^3\\ +2 a B^3 r^5+2 a B^2 r^3-8 a \alpha  B r-8 a B r+\alpha -3 \alpha  A B r^4+A B r^4-2 \alpha  A r^3-\alpha  A r^2+A r^2+3 \alpha  B^2 r^4-B^2 r^4-2 \alpha  B r^3+10 \alpha  B r^2\\
    +1)-2 a r (A+B) \left(-2 \alpha +9 \alpha  A B r^4-3 A B r^4+\alpha  A r^3+3 \alpha  A r^2+A r^2-3 \alpha  B^2 r^4+B^2 r^4+\alpha  B r^3-13 \alpha  B r^2+B r^2-2\right)].
\end{multline}
The expressions for matter density and pressure related to quark matter can be obtained as follows:
\begin{multline}
    \rho_q =\frac{3}{1-3 \alpha }\large[\frac{4 B_g}{3}+\frac{e^{-2 A r^2}}{2 \kappa  r^2} \large[-a r (A+B)(-8 \alpha +\alpha  A r^3+12 \alpha  A r^2+A r^3+4 A r^2+\alpha  B r^3-4 \alpha  B r^2+B r^3-12 B r^2-8)\\
    + e^{A r^2}(2 a \alpha  A^2 r^4+8 a \alpha  A^2 r^3+2 a A^2 r^4+4 a \alpha  A B r^4+4 a A B r^4-16 a A B r^3-16 a \alpha  A r-16 a A r+2 a \alpha  B^2 r^4-8 a \alpha  B^2 r^3+2 a B^2 r^4-\\16 a B^2 r^3-16 a \alpha  B r-16 a B r+2 \alpha -\alpha  A r^3-2 \alpha  A r^2-A r^3+2 A r^2-\alpha  B r^3+2 \alpha  B r^2-B r^3+6 B r^2+2)\\
    + e^{2 Ar^2}(-a \alpha  A^2 r^4+4 a \alpha  A^2 r^3-a A^2 r^4+4 a A^2 r^3-2 a \alpha  A B r^4+8 a \alpha  A B r^3-2 a A B r^4+8 a A B r^3+8 a \alpha  A r+8 a A r-a \alpha  B^2 r^4\\
    +4 a \alpha  B^2 r^3-a B^2 r^4+4 a B^2 r^3+8 a \alpha  B r+8 a B r-2 \alpha +\alpha  A r^3-2 \alpha  A r^2+A r^3-2 A r^2+\alpha  B r^3-2 \alpha  B r^2+B r^3-2 B r^2+2 \beta  \kappa  r^2-2)\large]\large],
\end{multline}
\begin{multline}
    p_q=\frac{1}{2 r^2 (\kappa -3 \alpha  \kappa )}[e^{-A r^2} (2 a \alpha  A^2 r^4+8 a \alpha  A^2 r^3+2 a A^2 r^4+4 a \alpha  A B r^4+4 a A B r^4-16 a A B r^3-16 a \alpha  A r-16 a A r\\
    +2 a \alpha  B^2 r^4-8 a \alpha  B^2 r^3+2 a B^2 r^4-16 a B^2 r^3-16 a \alpha  B r-16 a B r+2 \alpha -\alpha  A r^3-2 \alpha  A r^2-A r^3\\
    +2 A r^2-\alpha  B r^3+2 \alpha  B r^2-B r^3+6 B r^2+2)-a \alpha  A^2 r^4+4 a \alpha  A^2 r^3-a A^2 r^4+4 a A^2 r^3\\
    -2 a \alpha  A B r^4+8 a \alpha  A B r^3-a r (A+B) e^{-2 A r^2} (-8 \alpha +\alpha  A r^3+12 \alpha  A r^2+A r^3+4 A r^2+\alpha  B r^3-4 \alpha  B r^2+\\
    B r^3-12 B r^2-8)-2 a A B r^4+8 a A B r^3+8 a \alpha  A r+8 a A r-a \alpha  B^2 r^4+4 a \alpha  B^2 r^3-a B^2 r^4+4 a B^2 r^3+8 a \alpha  B r+8 a B r-2 \alpha\\
     +\alpha  A r^3-2 \alpha  A r^2+A r^3-2 A r^2+\alpha  B r^3-2 \alpha  B r^2+B r^3-2 B r^2+2 \kappa  r^2 (\beta +4 \alpha  B_g)-2].
\end{multline}

\section{Boundary Condition}\label{sec5}
It is crucial to maintain the continuity of space-time on both the inside and outside of the stellar model. The inner and exterior space times must thus coincide at the boundary $r=R$. Since we have taken static, non-rotating, and spherically symmetric spacetime, the exterior solution is represented by the following line element, which is a de-Sitter-Schwarzschild vacuum solution given as \cite{Bhar:2023bcd},
\begin{eqnarray}
ds_+^{2}&=&\left(1-\frac{2\mathfrak{m}}{r}+\frac{\Lambda r^2}{3}\right)dt^{2}-\left(1-\frac{2\mathfrak{m}}{r}+\frac{\Lambda r^2}{3}\right)^{-1}dr^{2}-r^{2}\left(d\theta^{2}+\sin^{2}\theta d\phi^{2}\right),
\end{eqnarray}
where `$\mathfrak{m}$' represents the total mass within the boundary of the compact star corresponding to the interior spacetime:
\begin{eqnarray}
ds_-^{2}&=&e^{Br^2+C}dt^{2}-e^{Ar^2}dr^{2}-r^{2}\left(d\theta^{2}+\sin^{2}\theta d\phi^{2}\right),
\end{eqnarray}
where the interior and exterior of spacetime are indicated, respectively, by the ($-$) and ($+$) signs.\\
Since the present value of the cosmological constant is given as $\Lambda=1.1056\times 10^{-46}~km^{-2}$, we can neglect it for our further consideration.\\
To match the interior space-time with the exterior at the boundary let us impose the continuity of $g_{rr},\,g_{tt}$ and $\frac{\partial}{\partial r}(g_{tt})$ across the boundary $r=R$ which gives the following set of equations. The above technique was used earlier in ref \cite{Deb:2018sgt,Rahaman:2010mr,Kalam:2012sh, Rahaman:2011hd}.
\begin{eqnarray}
1-\frac{2\mathfrak{m}}{R}&=&e^{B R^{2}+C},\label{u1}\\
\left(1-\frac{2\mathfrak{m}}{R}\right)^{-1}&=&e^{AR^{2}},\\
\frac{\mathfrak{m}}{R^{2}}&=&BRe^{B R^{2}+C}.\label{u3}
\end{eqnarray}
We solve the Equations (\ref{u1})-(\ref{u3}) simultaneously to obtain the expressions for $A,\, B,\,$ $C$ as,
\begin{eqnarray}
A&=&-\frac{1}{R^{2}}\ln\left(1-\frac{2\mathfrak{m}}{R}\right) ,\\
B&=&\frac{\mathfrak{m}}{R^{3}}\left(1-\frac{2\mathfrak{m}}{R}\right)^{-1} ,\\
C&=&\ln\left(1-\frac{2\mathfrak{m}}{R}\right)-\frac{\mathfrak{m}}{R}\left(1-\frac{2\mathfrak{m}}{R}\right)^{-1}.
\end{eqnarray}
According to O'Brien-Synge \cite{e1,e2}, for an isotropic fluid sphere, $p(R-0)=p(R+0)$, where $R$ denotes the boundary of the star. Since we are dealing with pressure anisotropy, for our present case we choose $p_r(R-0)=p_r(R+0)$. Since there is no matter outside the boundary of the star $ p_r(R+0)=0$, which gives,
\begin{eqnarray}
   \beta&=& 2\alpha \bigg[\frac{1}{R^2\kappa}
      e^{-2 A R^2} \Big[a (A + B) R \left\{-8 + R^2 (B (R-10) + A (6 + R))\right\} +
        e^{2 A R^2} \Big\{2 + (A + B) R (-(R-2) R \nonumber\\&&+ 
             a (-8 + (A + B) (R-4) R^2))\Big\} + 
       e^{A R^2} \Big\{-2 + 
          R \Big((B (R-5) + A (R-1)) R \nonumber\\&&+ 
             2 a (A + B) \big(8 - R^2 (A - 7 B + (A + B) R)\big)\Big)\Big\}\Big] - 
    2 B_g\bigg].
\end{eqnarray}
Table~\ref{tb8} represents the numerical values of the constants present in metric coefficients for different compact stars.
\begin{table*}[t]
\centering
\caption{ Values of $A,\,B$, and $C$ for different compact stars.}\label{tb8}
\begin{tabular}{@{}ccccccccccccc@{}}
\hline
Star & Observed mass & Observed radius & Estimated  & Estimated &  $A$&$B$&$C$\\
& $M_{\odot}$ & km. & mass ($M_{\odot}$) & radius (km.)& $km^{-2}$ & $km^{-2}$\\
\hline
EXO 1785-248 \cite{Ozel:2008kb} & $1.3 \pm 0.2$ & $8.85 \pm 0.4$ & 1.2 & 9.2 & 0.00573936 & 0.00369472 & $-0.7985$\\
Vela X-1 \cite{Rawls:2011jw} & $1.77 \pm 0.08$ & $9.56 \pm 0.08$    & 1.77 & 9.5& 0.01164 & 0.0102998 & $-1.98007$     \\
SMC X-4 \cite{Rawls:2011jw} & $1.29 \pm 0.05$ & $8.831 \pm 0.09$ & 1.29 & 8.8 & 0.00925394 & 0.00676338 &$-1.24038$ \\
LMC X-4 \cite{Rawls:2011jw} & $1.04 \pm 0.09$ & $8.301 \pm 0.2$ & 1.04 & 8.3 & 0.0083468& 0.00564051 & $-0.963586$  \\
4U 1538-52 \cite{Rawls:2011jw}& $0.87 \pm 0.07$ & $7.866 \pm 0.21$ & 0.87 & 7.8 & 0.00810485 &0.00523819&$-0.81179$  \\
\hline
\end{tabular}
\end{table*}

\section{Mass-radius curve}\label{sec6}
\begin{figure}[htbp]
    \centering
        \includegraphics[scale=.6]{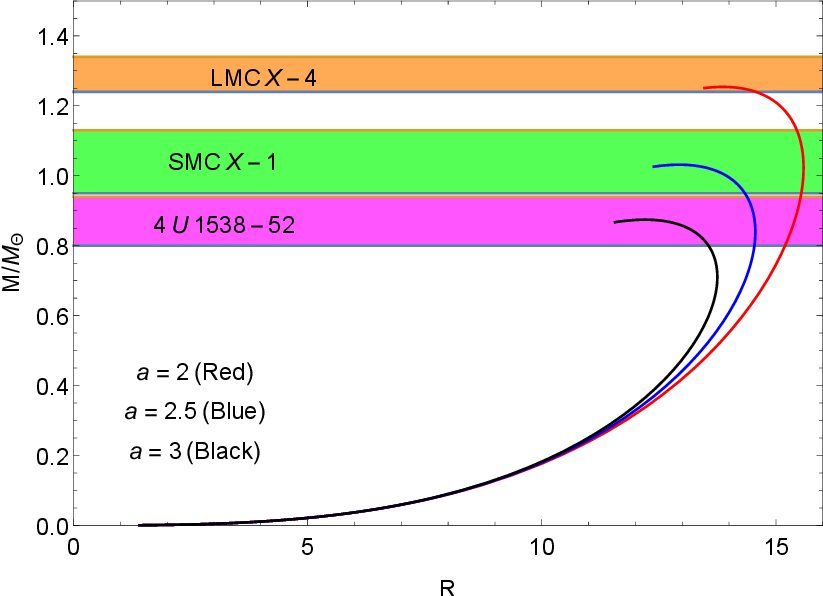}
       \caption{Mass-radius relationship are shown in the figure. For different values of `a', our model can generate the masses of some well-known compact objects.}
    \label{mr6}
\end{figure}

For three different values of $a$, namely $a=2,\, 2.5,\,$ and $3$, we provide the mass-radius curve for hybrid stars in the $f(Q)$ theory of gravitation in Fig.~\ref{mr6}.  From top to bottom, the three different color strips show the range of mass corresponding to the stars of LMC X-4, SMC X-1, and 4 U 1538-52.  From the picture, it can be seen that the radii of stars roughly range from $12.5$ to $14$ km. This radius range falls within the radii boundaries of compact stars predicted by observed X-ray binaries and the gravitational wave event known as GW 170817. Additionally, Fig.~\ref{mr6} depicts the variation of maximum mass in relation to the value of '$a$'. Less massive stars are discovered for greater '$a$'.  As '$a$' increases, the radii corresponding to the maximum masses decrease. Table~\ref{tb9} presents the maximum mass and corresponding radius for various values of `$a$' derived from Fig.~\ref{mr6}.

\begin{table*}[t]
\centering
\caption{For different values of `$a$', the maximum mass and associate radius}\label{tb9}
\begin{tabular}{@{}ccccccccccccc@{}}
\hline
$a$& Maximum mass $M(M_{\odot})$ && Associate radius (in km.) && Matched with the mass of the compact star \\
\hline
2& 1.25 &&       14.1 && LMC X-4 \cite{Rawls:2011jw} \\
2.5& 1.03 &&       13.3 && SMC X-1 \cite{Rawls:2011jw} \\
3& 0.87 &&     12.55 && 4U 1538-52 \cite{Rawls:2011jw}\\
\hline
\end{tabular}
\end{table*}

\section{Physical Analysis}\label{sec7}
We address the viability of compact stars in this section using physical characteristics such as metric components, energy density, pressures, and gradients of pressure, anisotropy, EoS, equilibrium conditions, stability, energy conditions, red-shift function, mass function, and adiabatic index. We select various values of $f(Q)$ parameter $a$ for the configuration of the graphical results, like $a=2$, $a=4$, $a=6$, $a=8$, and $a=10$.
\begin{figure}[htbp]
    \centering
        \includegraphics[scale=.6]{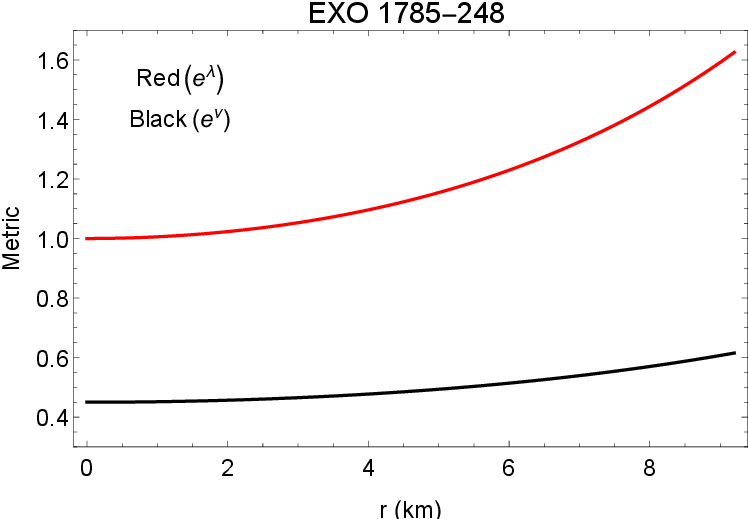}
    \caption{Both the metric potentials $e^{\nu}$ and $e^{\lambda}$ are drawn against `r'. The figure shows that both the metric potentials are monotonic increasing functions of `r' and are singularity-free.}
    \label{metric}
\end{figure}
\begin{enumerate}
\item For the solution to be realistic, energy density $\rho$ and both radial and transverse pressures $p_r$ and $p_t$ must remain positive throughout the interior of the fluid sphere, decrease monotonically as the radius grows, and be nonsingular inside the boundary. To verify the physical characteristics of pressures and energy density, we examine the graphical behavior of these material variables in Fig.~\ref{fig1}, which shows the positive nature of density and pressures in the interior of the star. Fig.~\ref{fig1} also confirms that there are no physical or geometric singularities in the current stellar system. However, the pressure anisotropy, which is quantified as $\Delta = p_t-p_r$ is positive when $p_t > p_r$ and negative otherwise, that is when $p_t<p_r$. Anisotropic factor also plays a significant role in the modeling of compact stars. Fig.~\ref{fig1} illustrates that the radial pressure in our current model is higher than the tangential pressure, which results in negative anisotropy. Negative anisotropy generates an attractive force that pulls inward.

\begin{figure}[htbp]
    \centering
        \includegraphics[scale=.4]{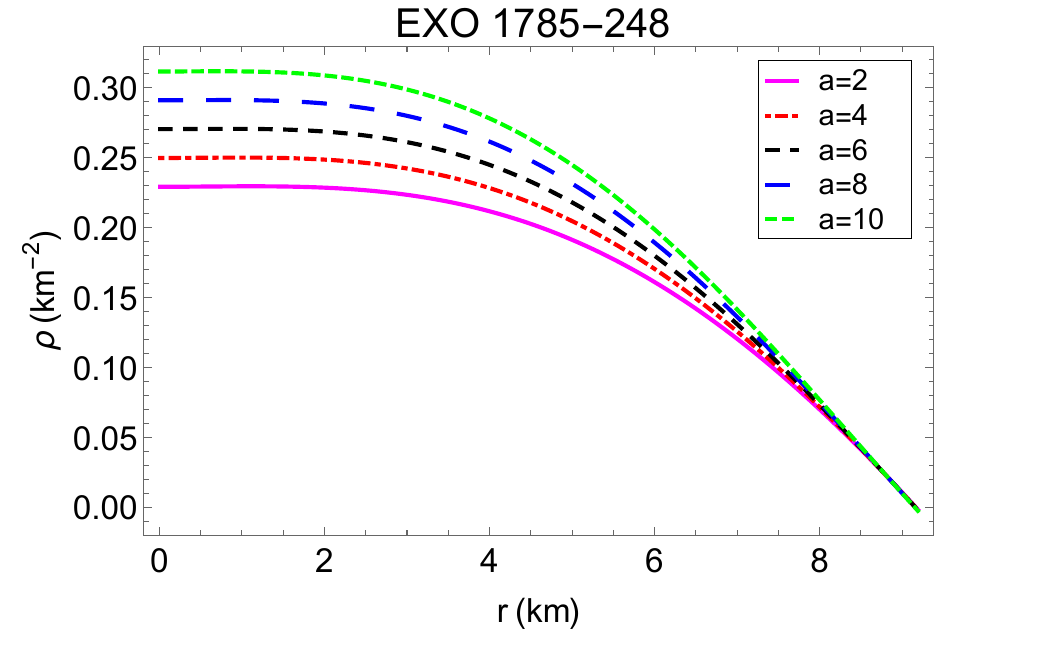}
        \includegraphics[scale=.4]{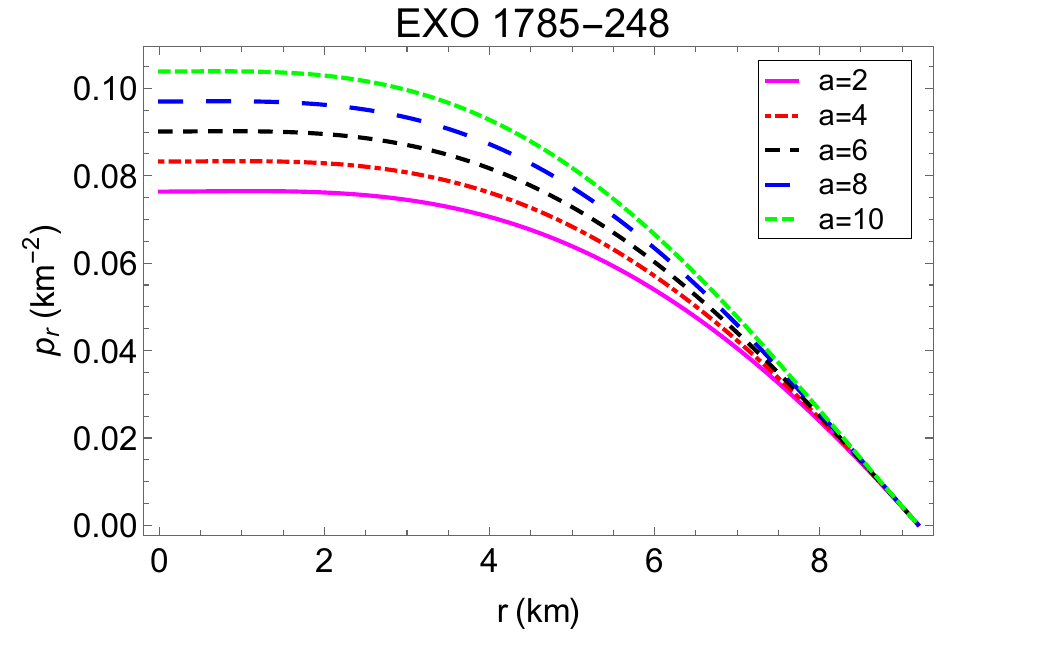}
        \includegraphics[scale=.4]{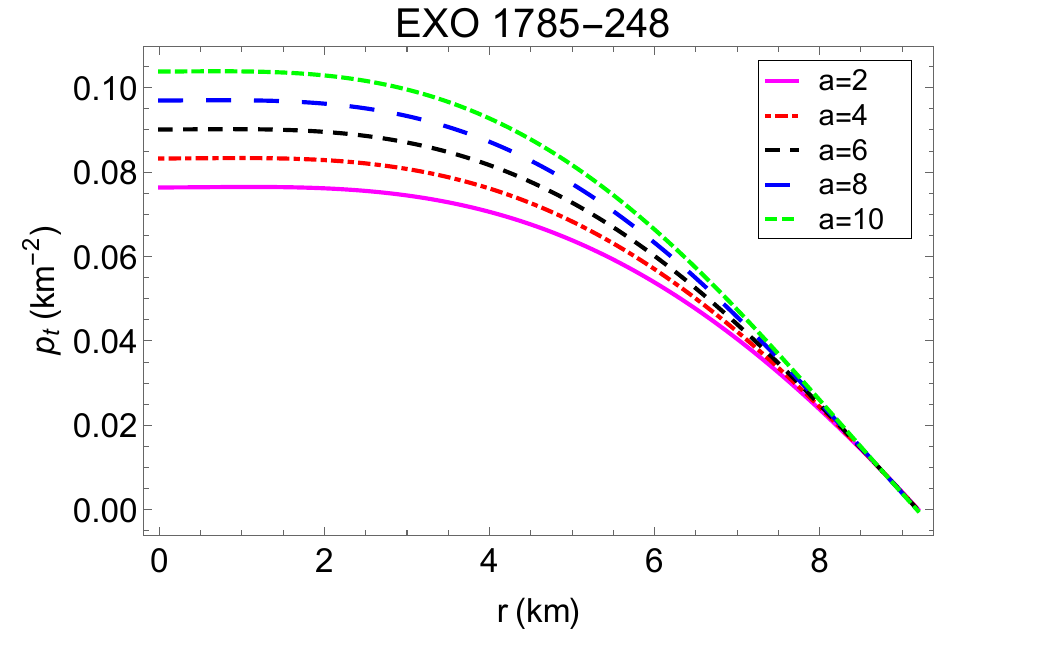}
        \includegraphics[scale=.4]{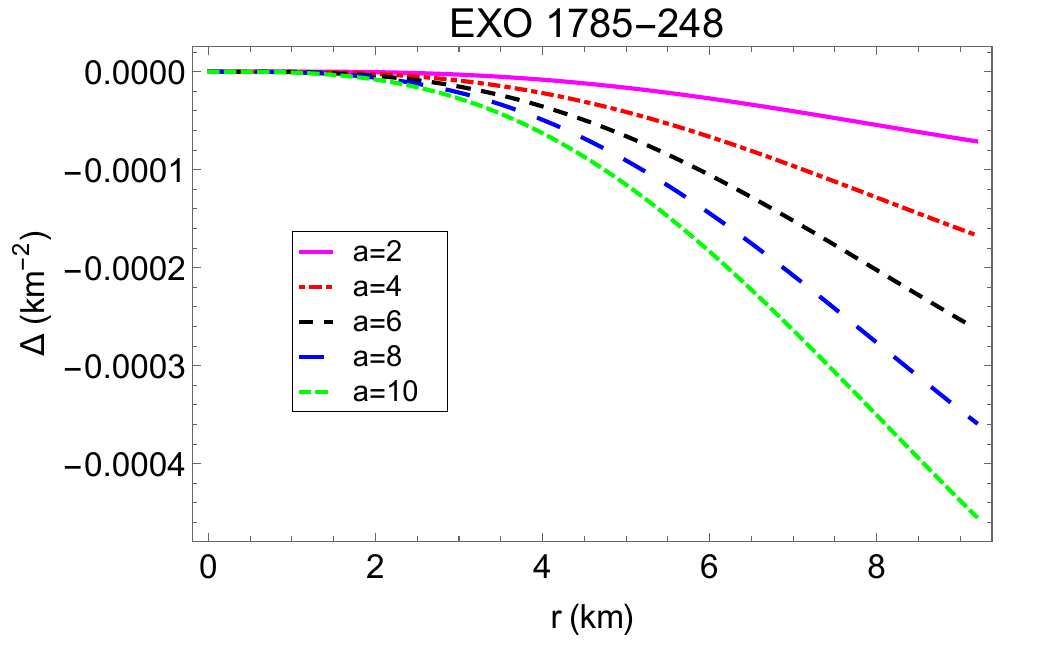}
 \caption{Matter density, radial pressure, transverse pressure, and anisotropic factor are shown for different values of $a$. The figure shows that density and both the pressures are monotonic decreasing functions of `$r$'. The anisotropic factor $\Delta<0$ implies that the underlying force is attractive.}
    \label{fig1}
\end{figure}

\item We draw the gradients of both the pressures and density inside the stellar interior in order to determine the optimal values for $\rho,\,p_r$ and $p_t$. We notice that the density and pressure gradients (Fig.~\ref{fig2}) are always negative throughout the interior and disappear at the center of the star, which confirms the fact that energy density and pressure are at their highest point at the core.
\begin{figure}[htbp]
    \centering
        \includegraphics[scale=.4]{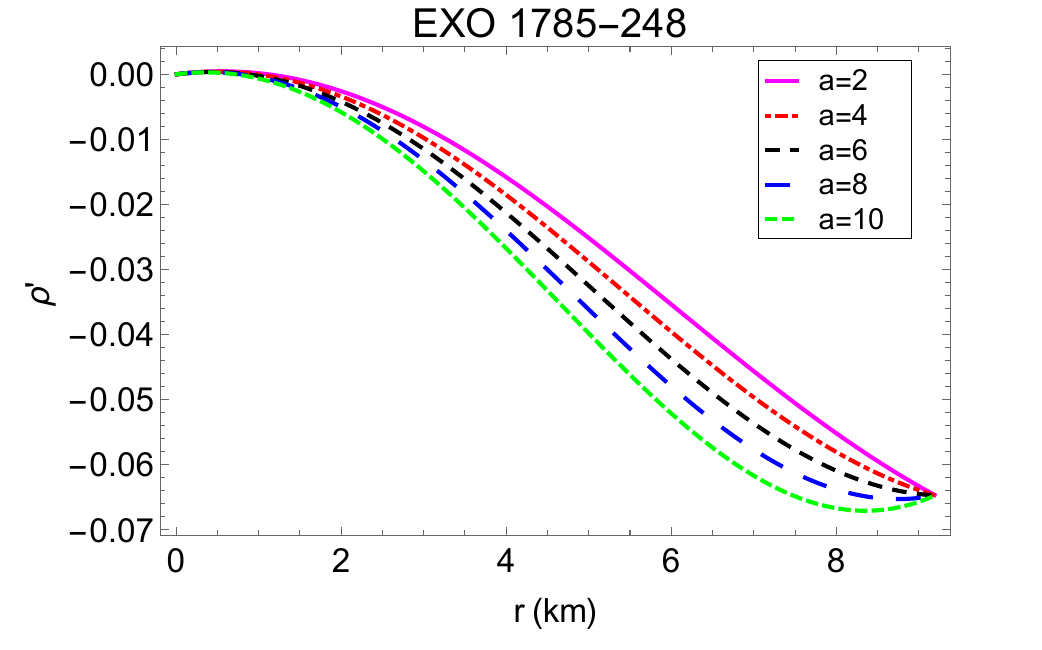}
        \includegraphics[scale=.4]{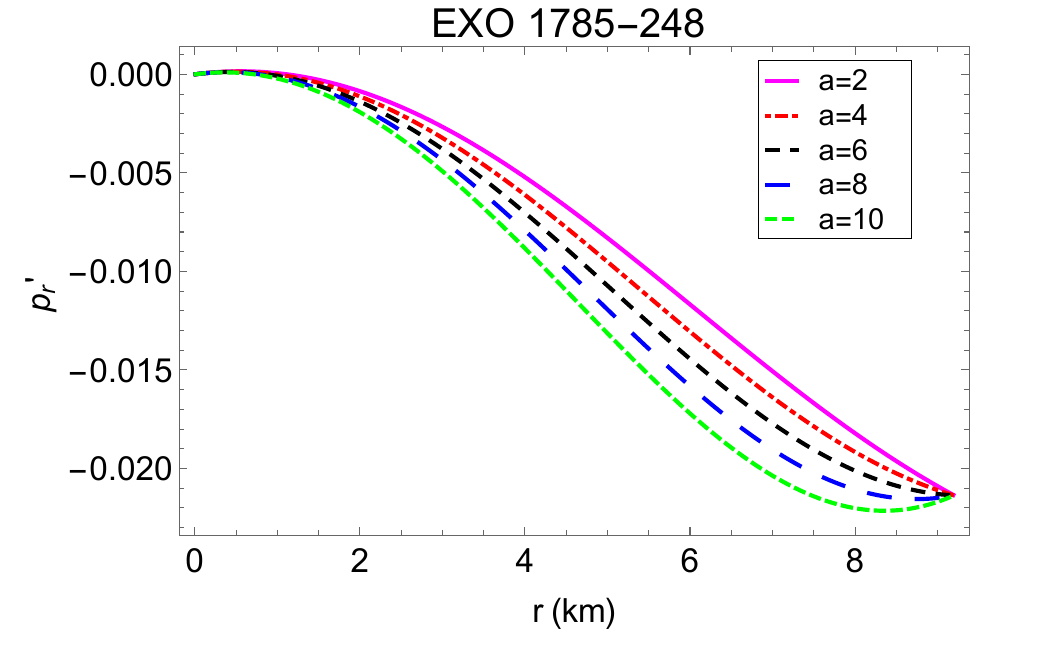}
        \includegraphics[scale=.4]{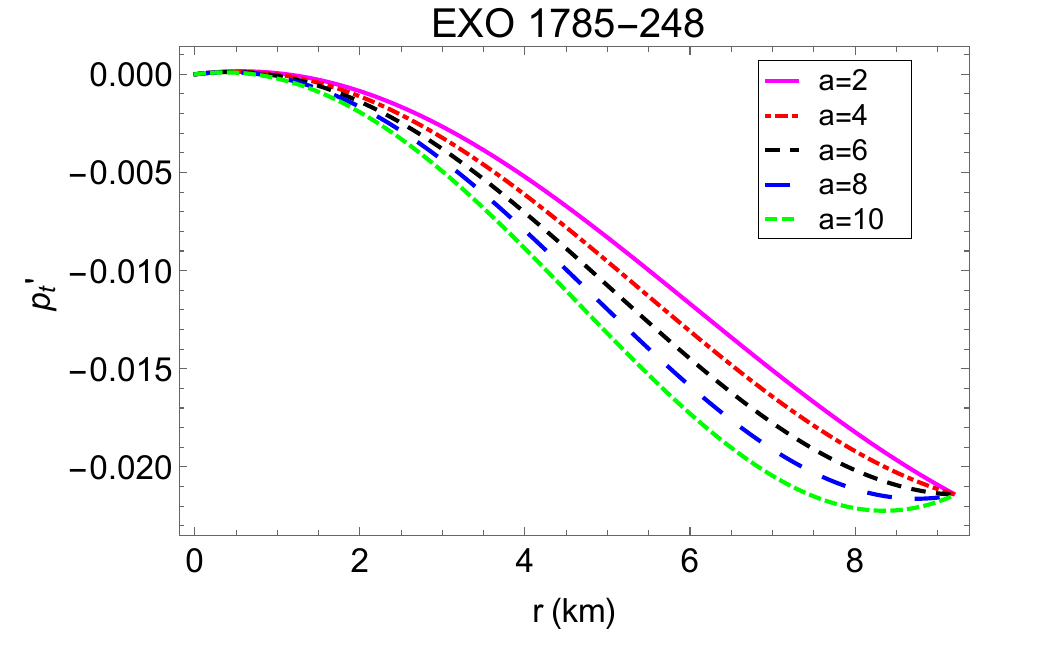}
       \caption{Density and pressure gradients are shown for different values of $a$. The negative values of pressure and density gradient verify that pressure and density have maximum values at their centre.}
    \label{fig2}
\end{figure}
\item In Fig.~\ref{fig3}, the graphical evolution of density and pressure $\rho_q$ and $p_q$ caused by quark matter is shown. Both quantities take a positive value for each of the coupling parameters '$a$' mentioned in the figure inside the stellar interior and monotonic decreasing towards the boundary.
\begin{figure}[htbp]
    \centering
        \includegraphics[scale=.4]{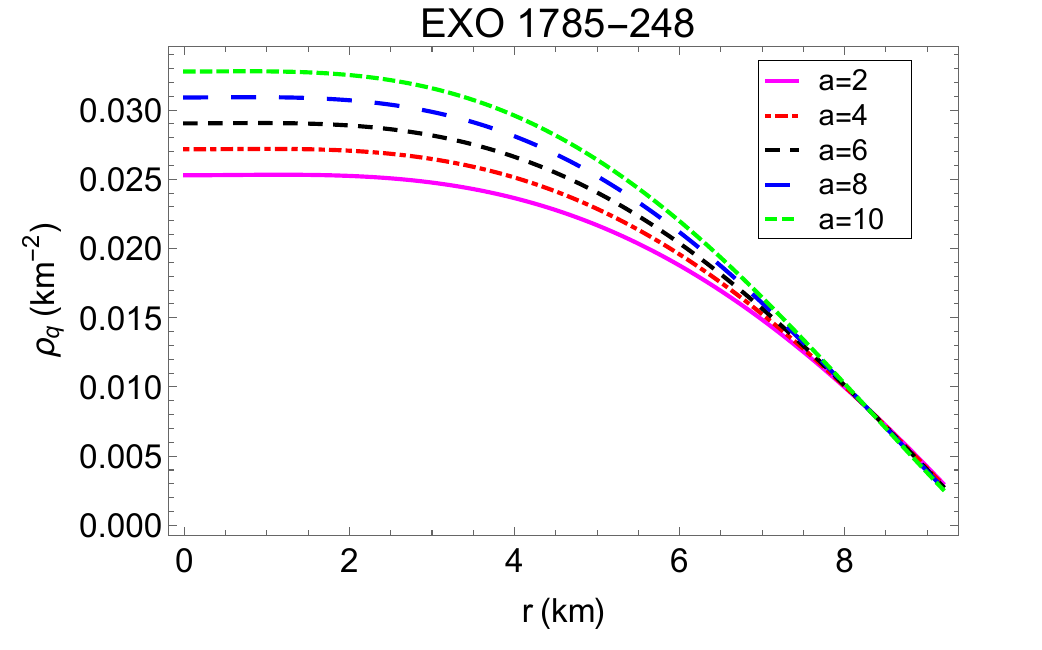}
        \includegraphics[scale=.4]{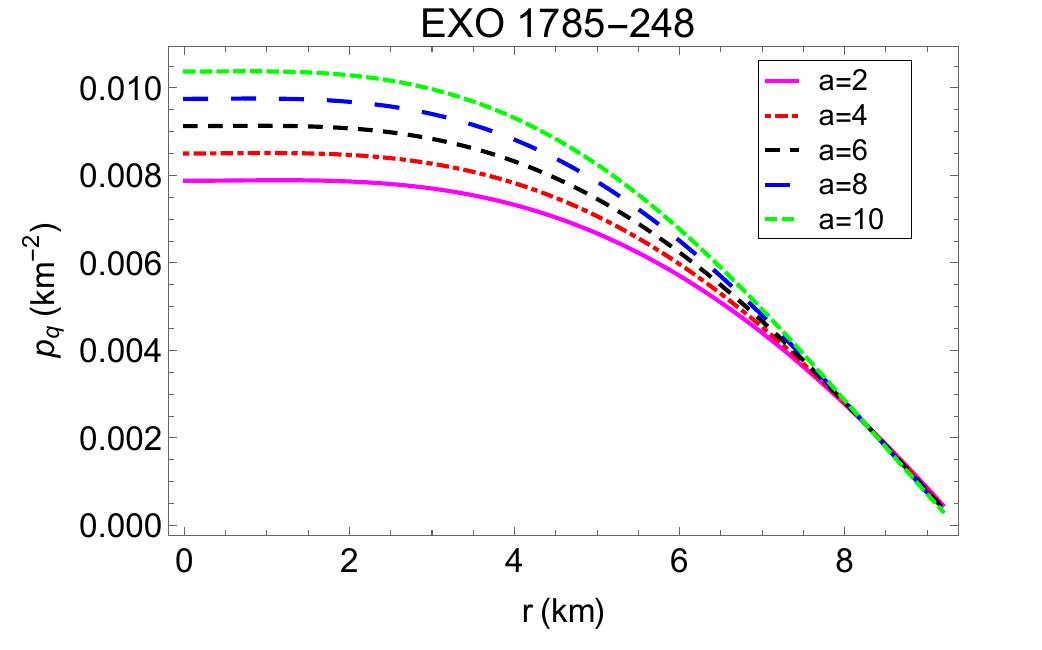}
       \caption{The behavior of density and pressure related to quark matter is shown for different values of $a$. The central values of density and pressure corresponding to quark matter increase with the increasing value of `a'.}
    \label{fig3}
\end{figure}
\item For our current model, the state parameters $\omega_r$ and $\omega_t$ are calculated as $\omega_r=p_r/\rho$ and $\omega_t=p_t/\rho$, respectively. Fig.~\ref{fig4} depicts the profiles of $\omega_r$ and $\omega_t$, which lies the limit ($0,\, 1$), demonstrating that the nature of the underlying matter in our system is not exotic in nature.
\begin{figure}[htbp]
    \centering
        \includegraphics[scale=.4]{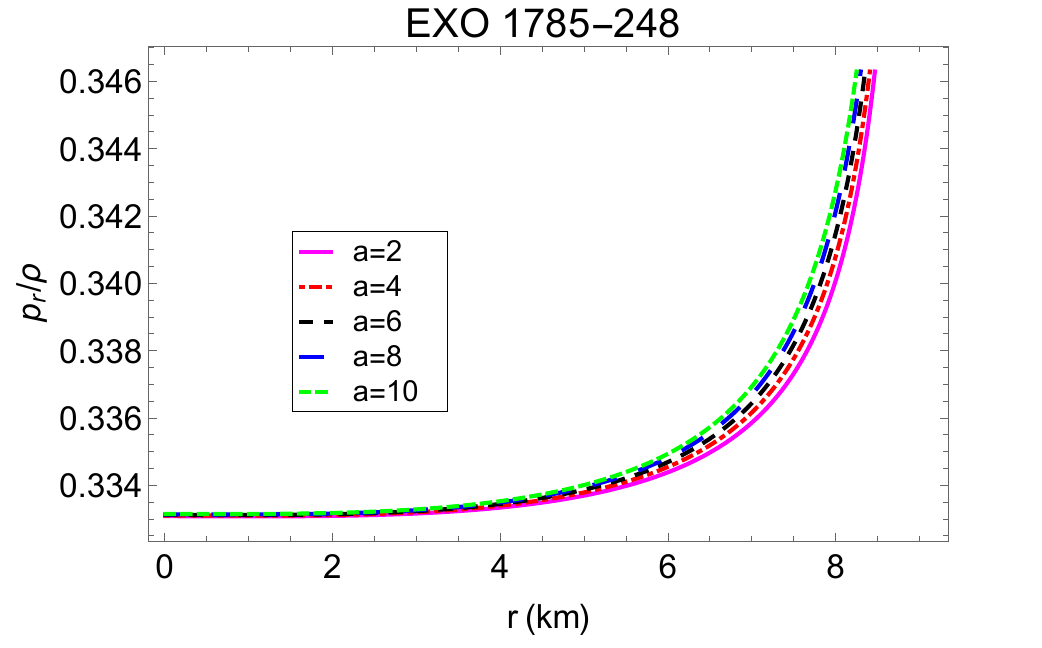}
        \includegraphics[scale=.4]{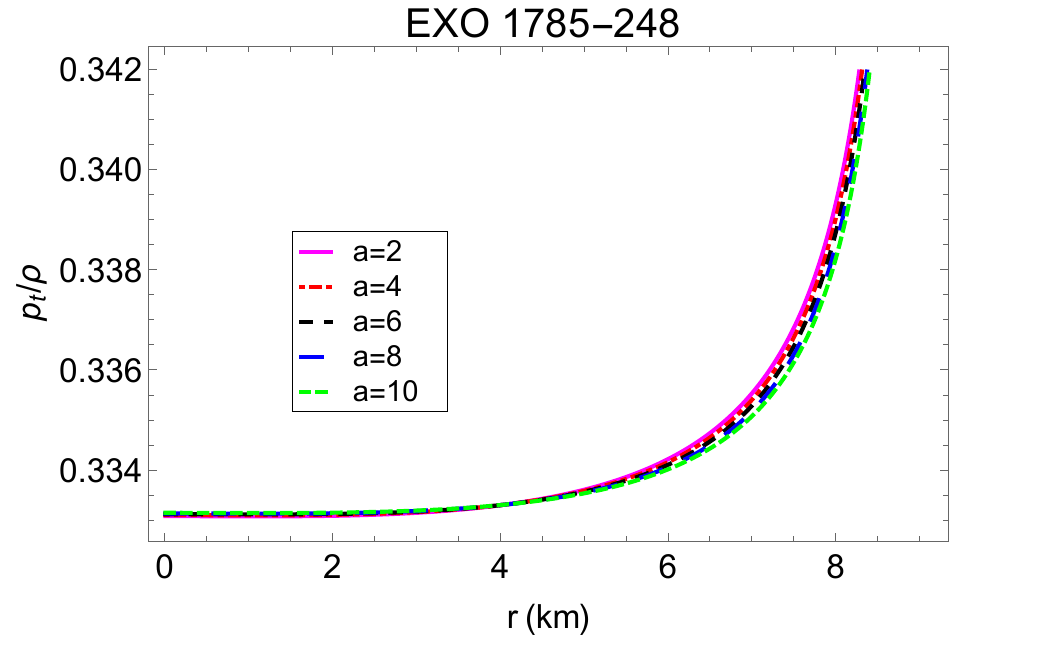}
       \caption{The ratio of pressure to density is shown for different values of $a$. The figure indicates that both the parameters lie in the range [$0,\,1$]indicating that the underlying fluid is non-exotic.}
    \label{fig4}
\end{figure}
\item Some conditions that depend on the relationships between energy density and pressure should be met throughout the interior of the star in order to guarantee that the matter distribution in the stellar interior is realistic. Those constraints are known as energy conditions, which are divided into four categories, namely strong energy conditions (SEC), null energy conditions (NEC), weak energy conditions (WEC), and dominant energy conditions (DEC). Furthermore, these conditions are regarded as an essential factor in determining whether the material inside the stellar structure is normal or exotic. Energy conditions are another important factor to consider when examining the second law of thermodynamics and the Hawking-Penrose singularity theory \cite{hp}. The following inequality must be simultaneously satisfied in order to allow these energy conditions.
\begin{itemize}
    \item NEC: $\rho+p_r \geq 0,\, \rho+p_t \geq 0,$
     \item WEC: $\rho+p_r \geq 0,\, \rho+p_t \geq 0,\,\rho \geq 0,$
 \item SEC: $\rho+p_r \geq 0,\, \rho+p_t \geq 0,\,\rho+p_r+2p_t \geq 0,$
\item DEC: $\rho-p_r \geq 0,\, \rho-p_t \geq 0,\,\rho \geq 0.$
\end{itemize}
We depicted the plots of the inequalities mentioned above in Fig.~\ref{ec11} to examine whether these are valid. We observed that these requirements are fulfilled over the entire fluid sphere, ensuring that the distribution of matter is non-exotic. 
\begin{figure}[htbp]
    \centering
        \includegraphics[scale=.4]{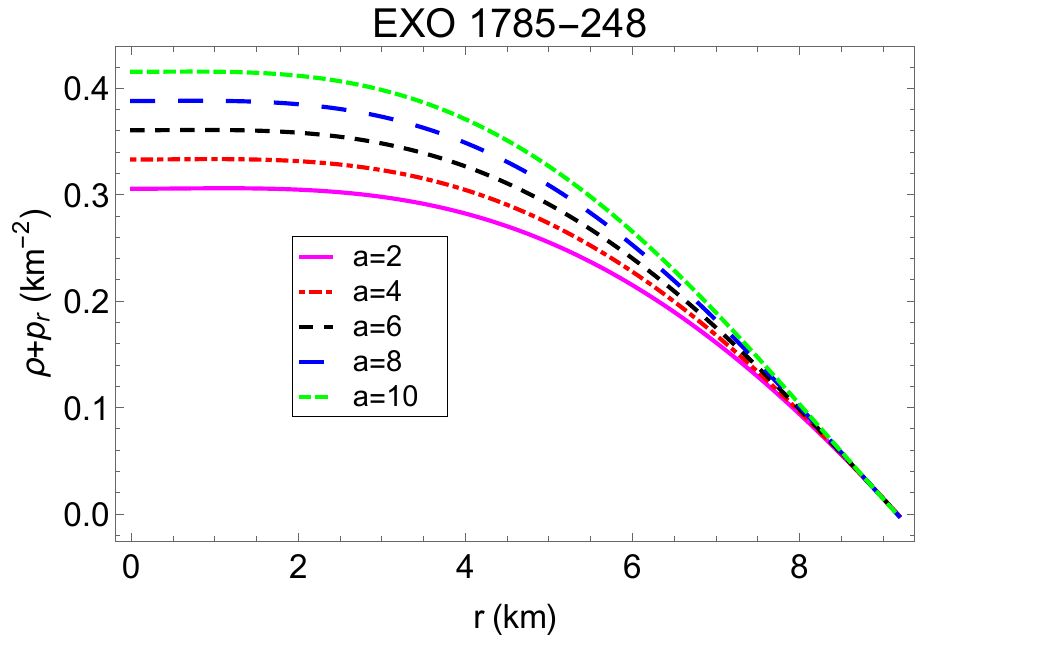}
        \includegraphics[scale=.4]{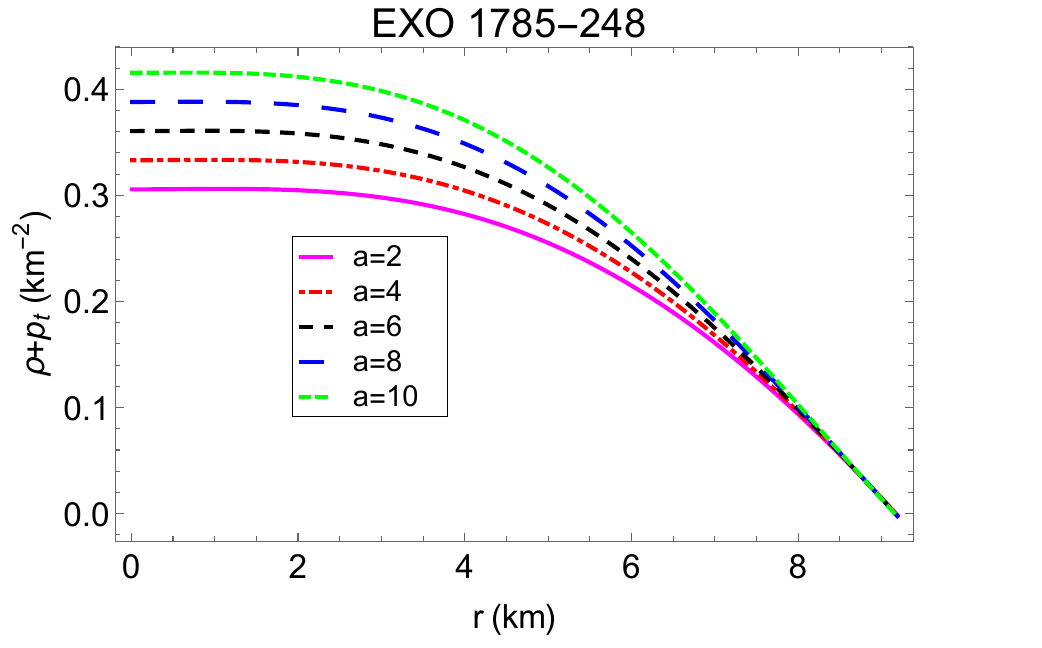}
        \includegraphics[scale=.4]{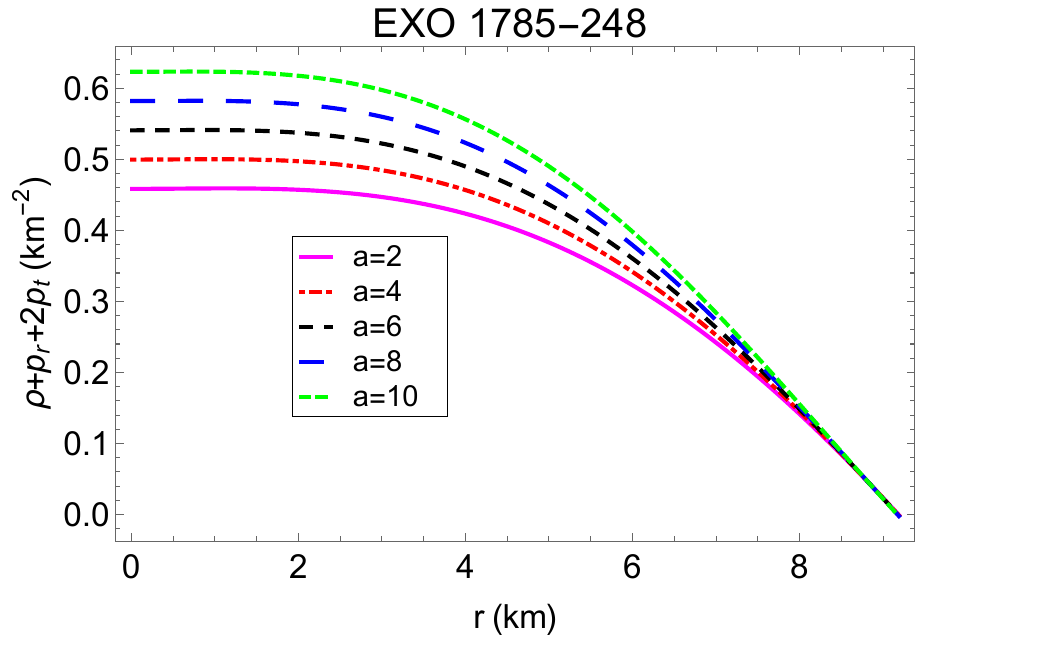}
        \includegraphics[scale=.4]{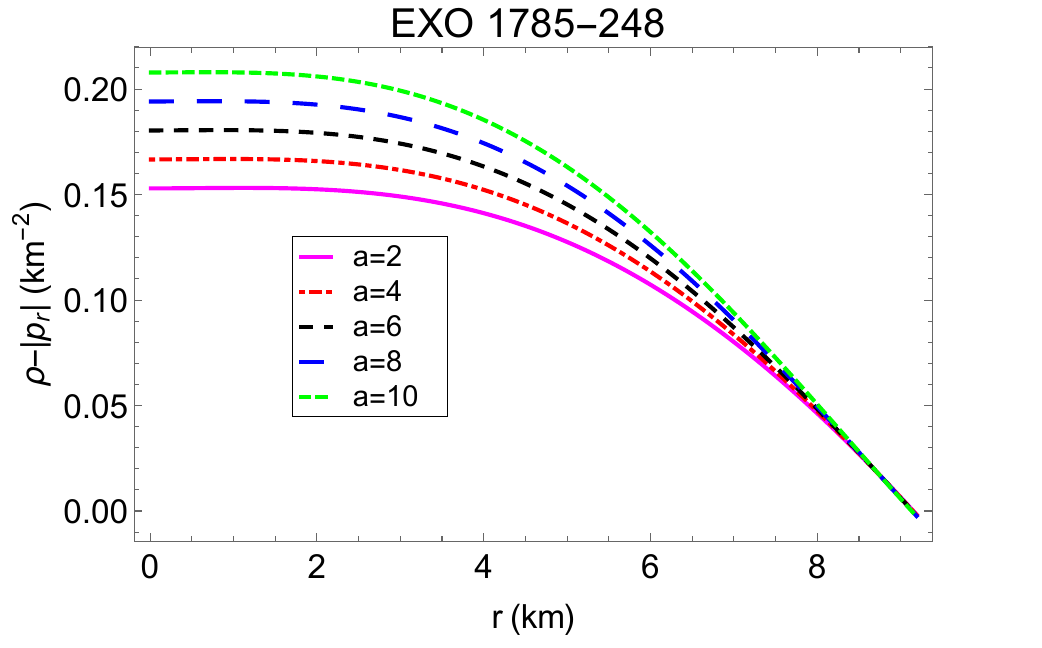}
        \includegraphics[scale=.4]{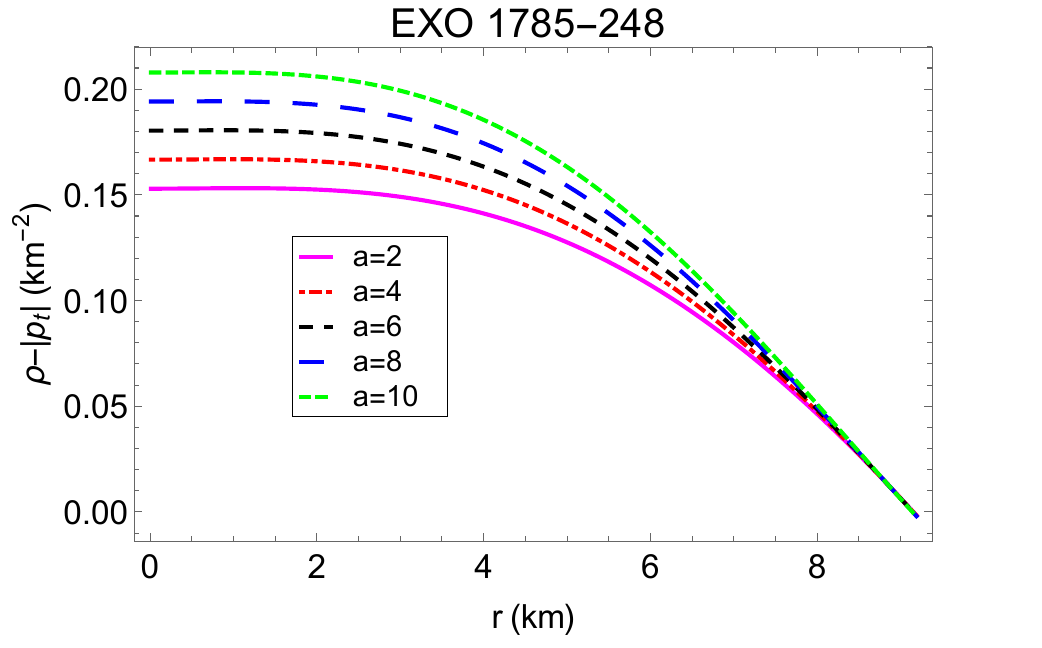}
\caption{The energy conditions are shown for different values of $a$. The figure indicates that all the energy conditions are well satisfied inside the fluid sphere for different values of `a'.}
    \label{ec11}
\end{figure}

\item The sound speed is a crucial parameter that needs to be restricted within a certain range to verify the model's stability. The square of the sound speed (tangential and radial) can be calculated as follows:
\begin{eqnarray}
    V_r^2=\frac{dp_r}{d \rho},\\
    V_t^2=\frac{dp_t}{d \rho}.
\end{eqnarray}
\begin{figure}[htbp]
    \centering
        \includegraphics[scale=.4]{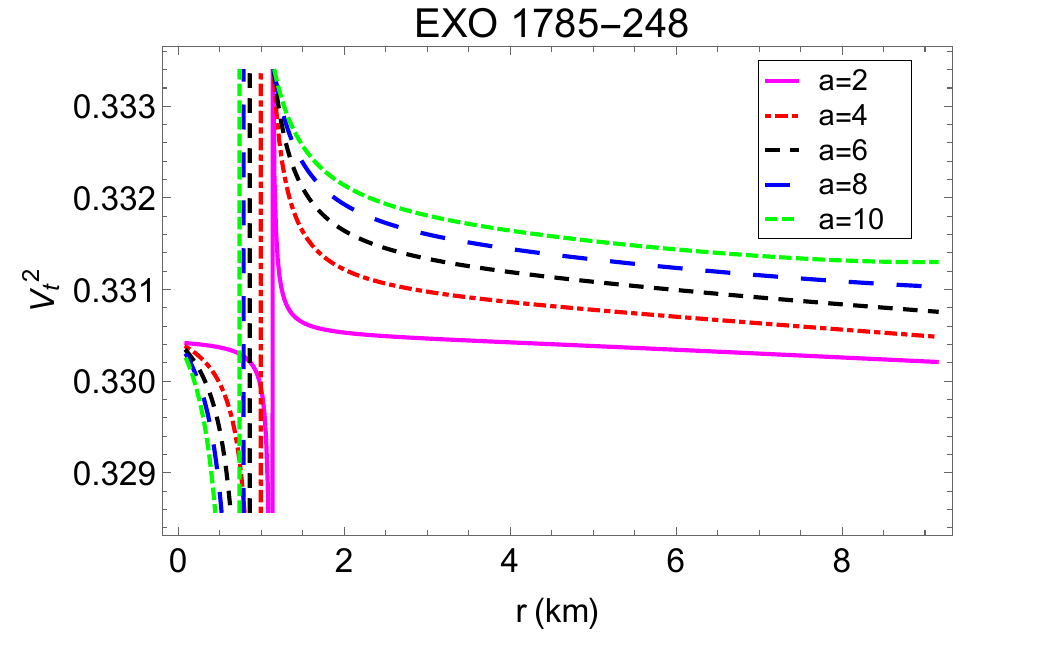}
        \includegraphics[scale=.4]{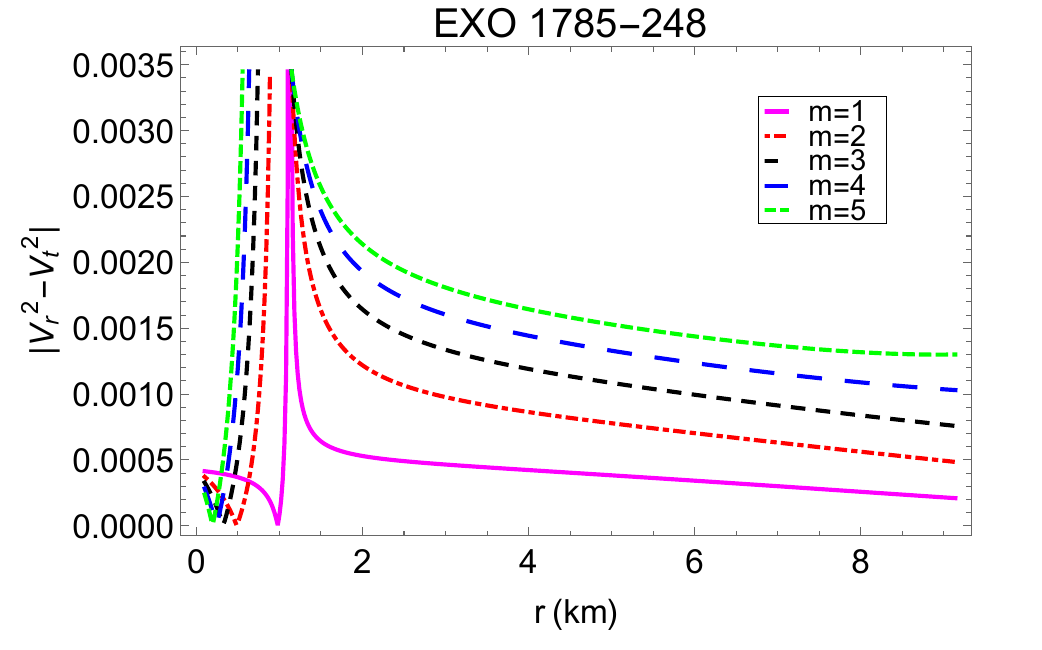}
\caption{Causality and stability conditions are shown for different values of $a$. The figure indicates that both the stability condition and the causality condition are well satisfied for different values of `a'.}
    \label{sv4}
\end{figure}

According to the causality condition, radial and transverse sound velocity must be obeyed for a stable model if the following inequalities hold inside the stellar interior:
$0\leq V_r^2 \leq 1$ and $0\leq V_t^2\leq 1$.
The square of the radial speed of sound is $\alpha$, which is taken as $0.33$ for drawing the plots, and therefore, it is less than $1$ for all different values of `$a$' for our present model.
The graph of $V_t^2$ is given in Fig.~\ref{sv4}, demonstrating that our suggested model satisfies the stability requirements outlined above.

\item A model that is vulnerable to cracking instabilities must be tested using both the causality condition and another crucial stability parameter, the cracking condition. The subliminal radial and tangential components of velocity are utilized to test the cracking instability in Herrera \cite{Herrera:1992lwz} and Abreu's \cite{Abreu:2007ew} criterion. Checking the cracking concept is crucial, especially in anisotropic fluid distributions. The cracking concept can be mathematically defined in terms of $|V_r^2-V_t^2|$. When the square of the radial speed of sound exceeds the square of the transverse speed of sound, a model could be considered stable. Andreasson \cite{Andreasson:2008xw} also claims that $|V_r^2-V_t^2|<1$ which is satisfied from its graphical illustration (Fig.~\ref{sv4}).

\begin{figure}[htbp]
    \centering
        \includegraphics[scale=.4]{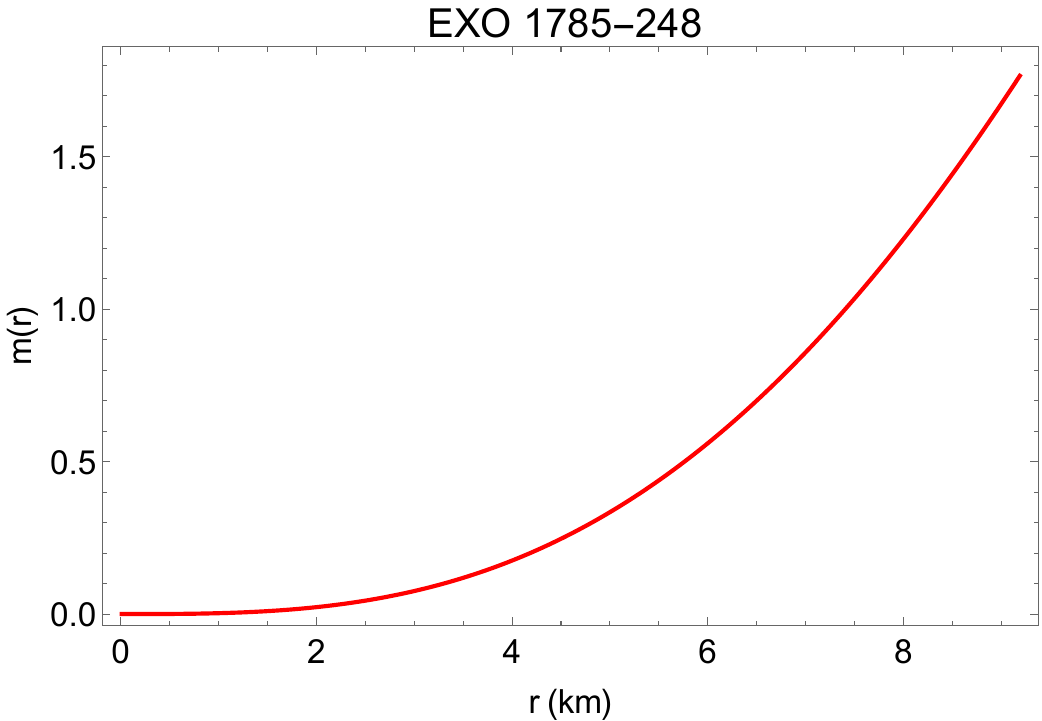}
        \includegraphics[scale=.4]{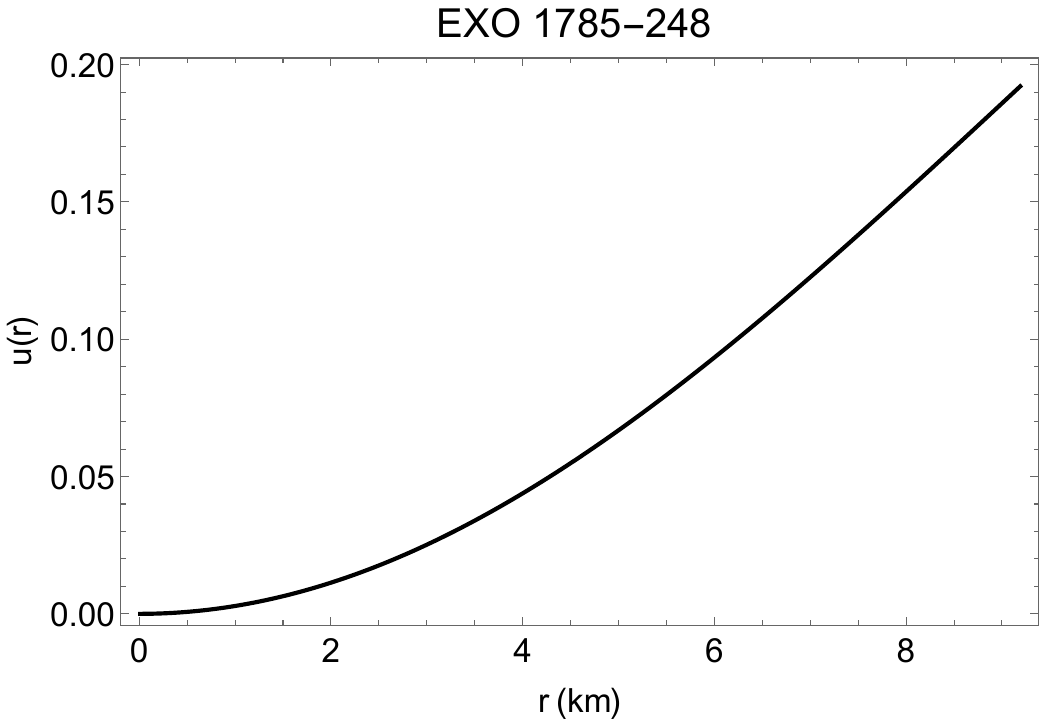}
        \includegraphics[scale=.4]{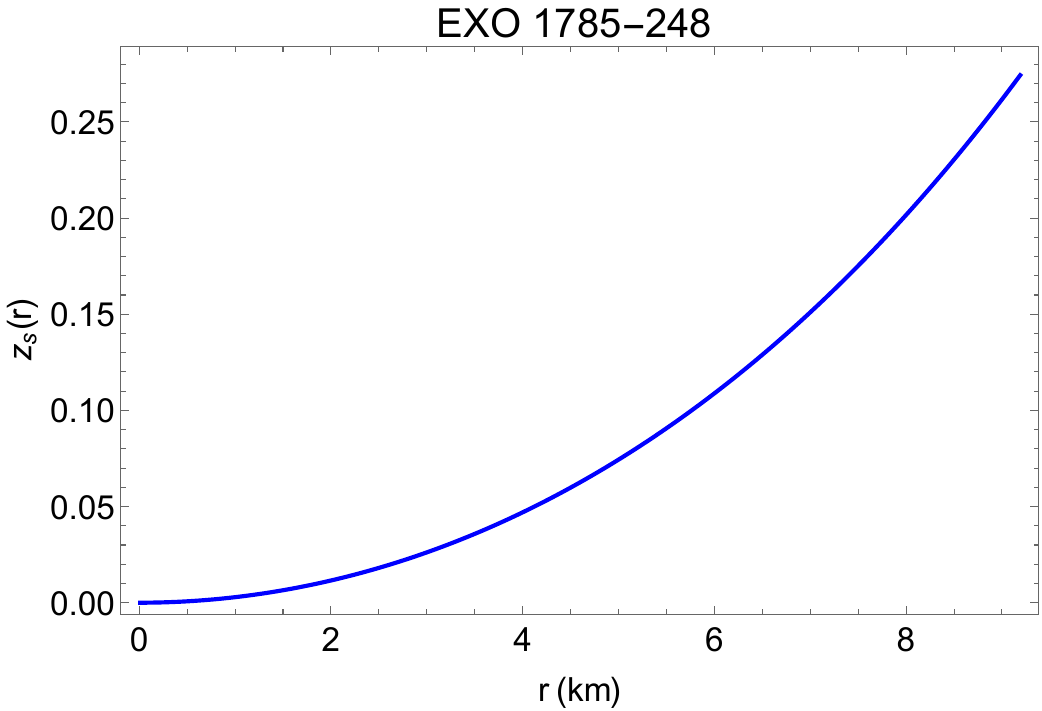}
       \caption{mass function, compactness, and surface redshift are shown inside the star. All the functions are monotonic increasing.}
    \label{fig8}
\end{figure}

\item For our model, the mass function is calculated as,
\begin{eqnarray}\label{n1}
M= 4\pi \int_0^R (\rho+\rho_q)r^2 dr,
\end{eqnarray}
where $R$ is the boundary of the star where the radius pressure vanishes. The equation (\ref{n1}) represents the baryonic mass of the star \cite{Carvalho:2022kxq}. The gravitational mass could be different from the baryonic one. The junction condition, i.e., $e^{-\lambda}=1-\frac{2m(r)}{r}$, can be used to determine the gravitational mass. The expression of the gravitational mass using the above condition can be obtained as, $m(r)=\frac{r}{2}(1-e^{-Ar^2})$.\\
The profile (Fig.~\ref{fig8}) of variation of mass $m(r)$ with respect to radius reveals that $m(r)$ vanishes at the center of the star, increasing monotonically from there. There are no singularities that affect the mass function. The formula $u(r)=m(r)/r$ yields the compactness factor of the model. Consequently, the surface redshift can be obtained as, $z_s(r)=(1-2u(r))^{-\frac{1}{2}}-1$. The nature of the compactness factor and surface redshift can be obtained from Fig.~\ref{fig8}. 
\begin{figure}[htbp]
    \centering
        \includegraphics[scale=.7]{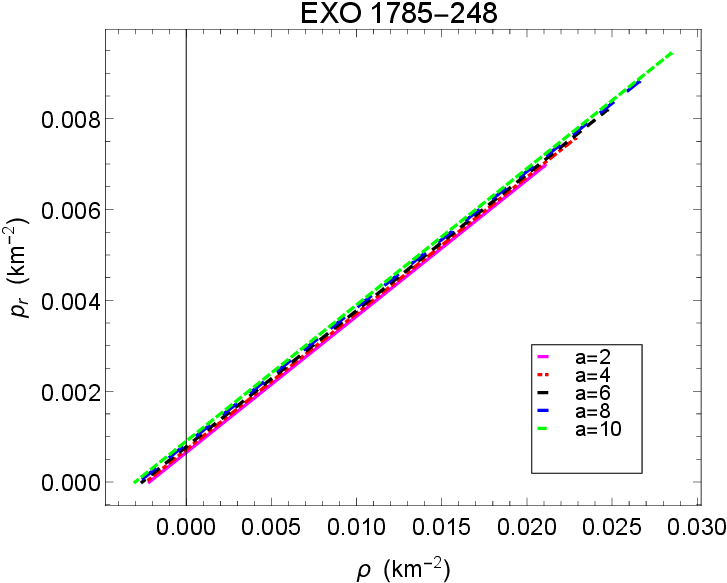}
        \includegraphics[scale=.5]{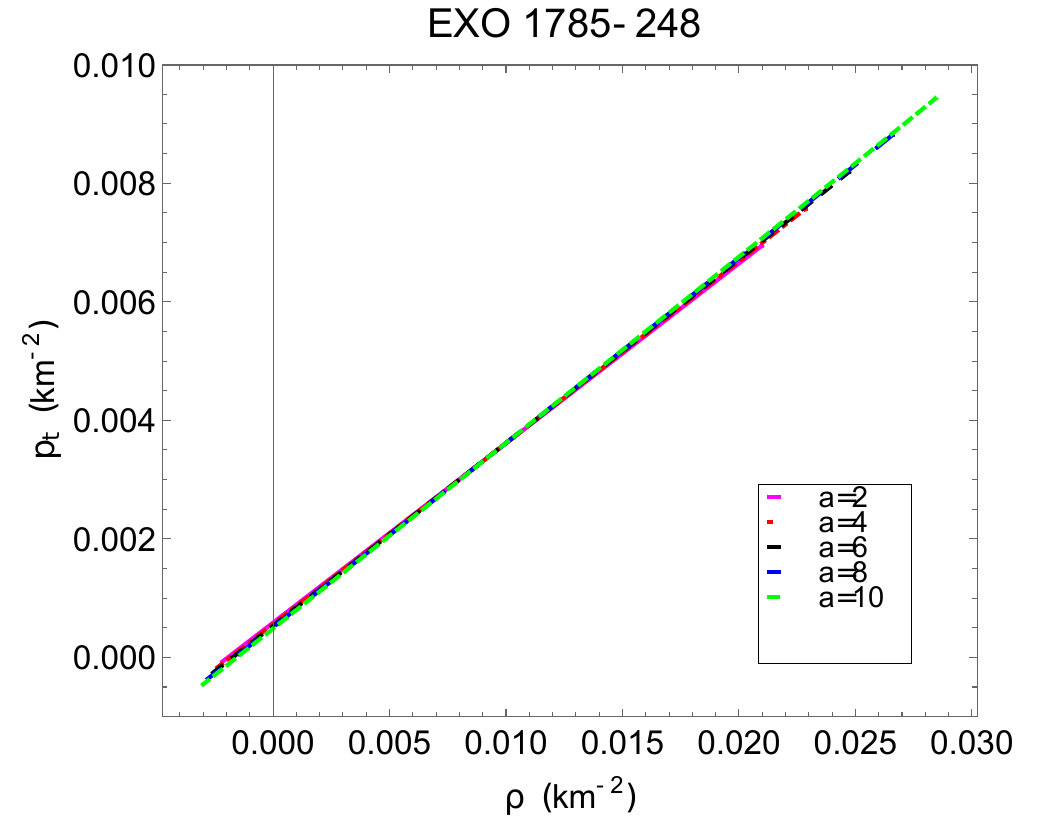}
       \caption{Variation of pressure with respect to density. The figure indicates that both the pressures maintain a linear equation of state with the density.}
    \label{par1}
\end{figure}

\item We used a graphical representation to check the fluctuation of both radial and transverse pressure with respect to density. According to Fig.~\ref{par1}, both pressures have a linear relationship with regard to density for different values of `$a$'.

\item 
Now we are interested in finding out the bound of the coupling parameter `a'.\\
Again, by applying the trace energy condition $\rho-p_r-2p_t~\geq 0$, at the surface of the star, we get the following bound for `a' as,
 \begin{eqnarray}
  a>\frac{1 + 5 B R^2 - R^3 (A + B + (A - B) B R) + 
 e^{A R^2} (-1 + (A + B) (-2 + R) R^2) + 
 2 e^{A R^2} R^2 \kappa B_g}{2 (A + B) R \left(4 + 8 B R^2 - (A + B) R^3 + 
   B (-A + B) R^4 + \Xi cosh(A R^2) + \eta \sinh(A R^2)\right)}=u(say),
   \end{eqnarray}
where $\Xi=\left\{-4 + 
      R^2 (A R + B (-8 + R + (3 A - B) R^2))\right\}$, $\eta= R^2 (-4 A + 4 B + B (-3 A + B) R^2)$.
      From the relationship that the transverse sound velocity is less than $1$ inside the stellar interior, i.e., $\frac{dp_t}{d\rho}<1$ at the boundary of the star gives us the following bound for `a'.
      \begin{eqnarray}
         a<1+\frac{e^{A R^2} (-1 - \alpha + 
   e^{A R^2} C_1 + 
   B R^3 C_2 + 
   A^2 R^4 C_3 - 
   A R^2 )}{ (A + B) R \left[2 (1 + \alpha) + 
   2 e^{2 A R^2} C_5 + 
   R^2 C_6 + 
   e^{A R^2} C_7\right]}=v (say),
      \end{eqnarray}
where the constants $C_i$'s are given as,
\begin{eqnarray*}
    C_1&=&1 + (1 + (A + B) R^3) \alpha,\\
    C_2&=&-\alpha + B R (-1 + 3 \alpha),\\
    C_3&=&-1 + \alpha + R (2 \alpha + B R (-1 + 3 \alpha)),\\
    C_4&=&1 + \alpha + 
      R (\alpha + 
         B R (-1 + 13 \alpha + 
            R (-2 \alpha+ B R (-1 + 3 \alpha)))),\\
            C_5&=&1 + (1 + (A + B) (-2 + R) R^2) \alpha,\\
            C_6&=&B (1 - 13 \alpha + 
         R (3 B R + 2 \alpha - 9 B R \alpha)) - 
      4 A^2 R^2 (1 + 3 \alpha + 
         R (-3 B R + \alpha + 9 B R \alpha)) \nonumber\\&&+ 
      A \Big\{9 + 11 \alpha + 
         R \Big(2 \alpha + 
            B R (-13 + 79 \alpha + 
               4 R \big(-\alpha + B R (-1 + 3 \alpha))\big)\Big)\Big\},\\
               C_7&=&-4 (1 + \alpha) + 
      R^2 \Big[B \big(-1 - 3 B R^2 + (17 + R (-4 + 9 B R)) \alpha\big) + 
         2 A^2 R^2 \Big(1 - \alpha + 
            R (2 \alpha + B R (-1 + 3 \alpha))\Big) \nonumber\\&&+ 
         A \Big\{-9 - 7 \alpha + 
            R \Big(-4 \alpha + 
               B R \big(5 - 43 \alpha + 
                  2 R (B R + 2 \alpha - 3 B R \alpha)\big)\Big)\Big\}\Big].
\end{eqnarray*}
Combining the above two inequalities for `a', we obtain a reasonable bound for `a' as,
$u<a<1+v$.
\end{enumerate}
\section{Discussion}\label{sec8}
By using two fluid distributions (SQM and regular baryonic matter) in $f(Q)$ gravity, we have looked at new models of anisotropic compact fluid spheres. There are no mathematical or geometrical singularities in our present models. For the compact star EXO 1785-248, the variations of several model parameters have been graphically represented in Figs. \ref{mr6}-\ref{par1} by using five distinct values of $a=2, 4, 6, 8$, and $10$. EXO 1785-248 was chosen because it was a good example of hybrid star candidates. We used the bag constant $B_g = 65~MeV/fm^3$ and the parametric value of $0.3$ for $\alpha$ throughout the investigation.\\
In Fig.~\ref{mr6}, we show the variation of $M$ with respect to $R$ for various values of '$a$'. We discovered that both $M_{max}$ and $R_{max}$ gradually decrease as the value of '$a$' rises. Additionally, it should be noted that the greatest value of the mass corresponded to the three well-known compact stars LMC X-4, SMC X-1, and 4U 1538-52 for various values of '$a$'. Fig.~\ref{metric} depicts the profile of the metric potentials and demonstrates that both metric potentials are not infinite at the center. It demonstrates that there aren't any physical or geometrical singularities in our system. Fig.~\ref{fig1} displays the fluctuations of $\rho$, $p_r$, $p_t$, and the anisotropic factor $\Delta$. The greatest density and pressure functions are found to be at the center of the system, and they fall monotonically to reach their minimum values at the surface, confirming the regularity of the obtained solutions. We discover that as the coupling parameter '$a$' increases, the densities and both radial and transverse pressures all take on greater values. Since the anisotropic component is negative, the anisotropic force is therefore attractive in nature. This is also clear from the fluctuation of the anisotropic stress for the various values of '$a$' seen in Fig. ~\ref{fig1}. We carry out some physical experiments, such as the energy conditions, causality conditions, and Andreasson criterion. We have demonstrated in Fig.~\ref{ec11} that our system is consistent under all energy conditions. Additionally, Fig.~\ref{sv4} shows that our system satisfies the Andreasson requirement as well as being consistent with the causality condition by concurrently satisfying all of the inequalities represented as $0<V_r^2,\,V_t^2<1$. Inside the stellar interior, the gradients of pressure and density for various values of '$a$' are depicted, the curves are negative in nature. Additionally, we describe the quark matter-related character of density and pressure using the graphical representation in Fig.~\ref{fig3}  and it can be seen that both the parameters took negative value inside the boundary. The change in pressures with respect to density suggests that the relationship between pressure and density is linear.  The mass function profile is monotonically growing from the center to the boundary and is free of singularities. The greatest value of the compactness factor is 0.192, which falls under the Buchdahl-proposed limit $u(R)<4/9$ \cite{buch}. The highest surface redshift at the same time is 0.275. When the cosmological constant is absent, the surface redshift ($z_s$) lies in the range $z_s\leq 2$, as we found in the literature \cite{buch,48,49}. On the other hand, Bohmer and Harko \cite{49} showed that the surface redshift for an anisotropic star obeys the condition $z_s \leq 5$ in the presence of the cosmological constant. Our suggested model abides by this constraint. In this respect, we also want to mention that when we are working with KB metric, we only consider the continuity of $g_{rr},\,g_{tt}$, and $\frac{\partial}{\partial r}(g_{tt})$ across the boundary $r=R$. The continuity of $\frac{\partial}{\partial r}g_{rr}$ is not considered because the simultaneous solutions of the previous three equations give the values of {$A,\,B$}, and $C$. The systems become inconsistent if we consider the continuity of $g_{rr},\,g_{tt},\,\frac{\partial}{\partial r}(g_{tt})$, and $\frac{\partial}{\partial r}(g_{rr})$ across the boundary. The same technique was used earlier in \cite{Deb:2018sgt,Rahaman:2010mr,Kalam:2012sh, Rahaman:2011hd}.     \par
In this connection, we can also mention that compact stars usually rotate and have angular momentum, such as neutron stars, and therefore the fully spherical symmetric solution may not be applicable. But, for simplicity, we have considered non-rotating compact stars in our present paper and hence, we have not considered any angular momentum and several works have been done earlier in f(Q) gravity by considering the compact stars without rotation \cite{Bhar:2023bcd,Bhar:2024vxk,Bhar:2023zwi, Bhar:2023xku,Kaur:2024biz, Maurya:2024rlm, Maurya:2024dwn}.
Within the context of the $f(Q)$ theory of gravity, we investigated the KB-type metric using two fluid distributions (SQM and regular baryonic matter). By analyzing the compact star candidate in $f(Q)$ gravity, we reached a result of our inquiry that is extremely capable of demonstrating both the viability of the central singularity and the stability conditions for the internal fluid distribution. As a result, we can state that we anticipated a very dense, compact object consisting of quark matter that would be ideal for hybrid star objects for investigating their numerous physical features at both theoretical and astrophysical scales. Additionally, we note that, in addition to the conventional GR approach, other higher-order gravity theories can also be used to generate these physical properties under some specified constraints.

In the study of hybrid stars, modified gravity theories such as $f(R)$, $f(T)$, and $f(Q)$ provide a variety of frameworks for understanding stellar structures and their mysterious traits. Each theory has distinct features that influence the modelling of hybrid stars, which are made up of both strange quark matter (SQM) and ordinary baryonic matter (OBM).
In $f(R)$ theory, GR is modified by introducing a function $f(R)$ that is dependent on the Ricci scalar R. It has been extensively researched and can explain a variety of cosmic issues, including dark energy and the dynamics of compact stars \cite{hr1,hr01}. However, it frequently suffers with singularities in stellar models and may not fully represent some exotic matter states \cite{hr02,hr03}.
Teleparallel gravity $f(T)$ is based on torsion rather than the curvature. This allows for distinct gravitational dynamics and gives interesting insights into cosmology and gravitational waves, and large-scale structures \cite{hr2}. Its applications to hybrid stars produces insightfull results \cite{hr3}. The newer $f(Q)$ theory accentuates nonmetricity Q as a key component of gravity. The $f(Q)$ gravity theory is considered as a more flexible technique to modelling gravitational interactions, especially in the environments with anisotropic pressures and non-singular star configurations. $f(Q)$ gravity may support non-singular hybrid stars, resulting in smooth interior structures. Studies \cite{r43,Bhar:2023bcd,Bhar:2024vxk,Bhar:2023zwi, Bhar:2023xku,Kaur:2024biz, Maurya:2024rlm, Maurya:2024dwn} show that $f(Q)$ gravity can create supermassive hybrid stars, explaining the observed mass-radius relationship in compact stars. This modeling flexibility is critical for integrating theoretical predictions with empirical evidence. The framework addresses anisotropic pressures in stellar interiors, which is crucial for adequately modeling hybrid star compositions.
$f(Q)$ is a new theory with limited empirical support compared to established theories like $f(R)$. Its consequences are currently being investigated, which may contribute to uncertainty in projections. Comparative investigations are necessary to properly grasp the advantages and limitations of this theory in different astrophysical environments, notwithstanding initial promising results.

In conclusion, while $f(Q)$ gravity offers novel techniques to modeling hybrid stars with non-singular solutions and a strong foundation for anisotropic pressures, it must be verified against established theories like as $f(R)$ and $f(T)$. As study develops, the distinct qualities of $f(Q)$ may either reinforce its position as a viable option or show flaws that require additional improvement. In the present study, our results in $f(Q)$ theory are well consistent with the observational evidence of the mass and radius of considered star and in good concurrence with earlier studies in $f(T)$ and $f(R)$ gravity theories.

\section*{Data availability} No new data is associated with this article.

\section*{Acknowledgements} P.B. is thankful to the Inter-University Centre for Astronomy and Astrophysics (IUCAA), Pune, Government of India, for providing a visiting associateship.  SM gratefully acknowledges the Japan Society for the Promotion of Science (JSPS) for providing the postdoctoral fellowship for 2024-2026 (JSPS ID No.: P24026). This work of SM is supported by the JSPS KAKENHI Grant (Number: 24KF0100). PKS acknowledges the National Board for Higher Mathematics (NBHM) under the Department of Atomic Energy (DAE), Govt. of India, for financial support to carry out the Research project No.: 02011/3/2022 NBHM(R.P.)/R\&D II/2152 Dt.14.02.2022 and IUCAA, Pune, India for providing support through the visiting Associateship program.
We are very much grateful to the honorable referee and to the editor for the illuminating suggestions that have significantly improved our work in terms of research quality, and presentation.



\begin{thebibliography}{64}
\bibitem{hp}S.W. Hawking, G.F.R. Ellis, The Large Scale Structure of Space-Time (Cambridge University Press,
Cambridge, 1975)

\bibitem{r1} S. Perlmutter et al.,  Measurements of $\Omega$ and $\Large$ from 42 High-Redshift Supernovae, ApJ \textbf{517}, 565 (1999)
\bibitem{r2} C.L. Bennett et al., First-Year Wilkinson Microwave Anisotropy Probe (WMAP)* Observations: Preliminary Maps and Basic Results, Astrophys. J. Suppl. \textbf{148}, 1 (2003)
\bibitem{r3} A.G. Riess et al., Supernova search team collaboration. Observational Evidence from Supernovae for an Accelerating Universe and a Cosmological Constant, Astron. J. \textbf{116}, 1009 (1998)
\bibitem{r4} P.A.R. Ade et al., Detection of 
B-Mode Polarization at Degree Angular Scales by BICEP2, Phys. Rev. Lett. \textbf{112}, 241101 (2014)
\bibitem{r5}  W.M. Wood-Vasey et al., Observational Constraints on the Nature of Dark Energy: First Cosmological Results from the ESSENCE Supernova Survey, Astrophys. J. \textbf{666}, 694 (2007)
\bibitem{r6} M. Kowalski et al., Improved Cosmological Constraints from New, Old, and Combined Supernova Data Sets. Astrophys. J. \textbf{686}, 749 (2008)
\bibitem{r7} E. Komatsu et al., Five-year Wilkinson Microwave Anisotropy Probe Observations: Cosmological interpretation. Astrophys. J. Suppl. \textbf{180}, 330 (2009)
\bibitem{r8}  M. Tegmark et al., Cosmological parameters from SDSS and WMAP. Phys. Rev. D \textbf{69}, 103501 (2004)
\bibitem{r9} K. Abazajian et al., The Third Data Release of the Sloan Digital Sky Survey, Astron. J. \textbf{129}, 1755 (2005)
\bibitem{r10} K. Abazajian et al., The Second Data Release of the Sloan Digital Sky Survey. Astron. J. \textbf{128}, 502 (2004)
\bibitem{r11}  K. Abazajian et al., The First data release of the Sloan Digital Sky Survey. Astron. J. \textbf{126}, 2081 (2003)
\bibitem{r12} E. Hawkins et al., The 2dF Galaxy Redshift Survey: correlation functions, peculiar velocities and the matter density of the Universe, Mon. Not. R. Astron. Soc. \textbf{346}, 78 (2003)
\bibitem{r13} L. Verde et al.,The 2dF Galaxy Redshift Survey: the bias of galaxies and the density of the Universe , Mon. Not. R. Astron. Soc. \textbf{335}, 432 (2002)
\bibitem{r14} D.N. Spergel et al., First-Year Wilkinson Microwave Anisotropy Probe (WMAP)* Observations: Determination of Cosmological Parameters. ApJS \textbf{148}, 175 (2003)
\bibitem{r15} J. Edmund et al.,Dynamics of dark energy, Int. J. Mod. Phys. D \textbf{15}, 1753 (2006)
\bibitem{r16} S.Weinberg,The cosmological constant problem, Rev. Mod. Phys. \textbf{61}, 1 (1989).
\bibitem{r17} B. Ratra and P.J.E. Peebles, Cosmological consequences of a rolling homogeneous scalar field, Phys. Rev. D \textbf{37}, 3406 (1998).
\bibitem{r18} C. Armendariz-Picon et al., Dynamical Solution to the Problem of a Small Cosmological Constant and Late-Time Cosmic Acceleration, Phys. Rev. Lett. \textbf{85}, 4438 (2000).
\bibitem{r19} M. Sami and A. Toporensky, Phantom Field and the Fate of Universe, Mod. Phys. Lett. A \textbf{19}, 1509 (2004).
\bibitem{r20} J. Khoury and A. Weltman, Chameleon Fields: Awaiting Surprises for Tests of Gravity in Space, Phys. Rev. Lett. \textbf{93}, 171104
\bibitem{r21} M. C. Bento et al.,Generalized Chaplygin gas, accelerated expansion, and dark-energy-matter unification, Phys. Rev. D \textbf{66}, 043507 (2002).
\bibitem{r22} R. Zarrouki and M. Bennai, Chaplygin gas braneworld inflation according to WMAP7 data, Phys. Rev. D \textbf{82}, 123506 (2010).
\bibitem{r23} T. Padmanabhan, Accelerated expansion of the universe driven by tachyonic matter, Phys. Rev. D \textbf{66}, 021301 (2002).
\bibitem{oa1} A. Rincon, A Ovgun, .R.C. Pantig, An effective model for the quantum Schwarzschild black hole: Weak deflection angle, quasinormal modes and bounding of greybody factor, Physics of the Dark Universe, \textbf{46}, 101623 (2024).
\bibitem{oa2} G. Lambiase, R.C. Pantig, AOvgun, Traces of quantum gravitational correction at third-order curvature through the black hole shadow and particle deflection at the weak field limit, Physics of the Dark Universe, \textbf{46}, 101597 (2024).
\bibitem{oa3}A Ali, A Ovgun, Topological dyonic black holes of massive gravity with generalized quasitopological electromagnetism, The European Physical Journal C \textbf{84} (4), 378 (2024).
\bibitem{oa4} E Sucu, A Ovgun, The effect of quark-antiquark confinement on the deflection angle by the NED black hole, Physics of the Dark Universe \textbf{44}, 101446 (2024).
\bibitem{r23a} A. Qadir, H.W. Lee, K.Y. Kim, Motion of test particles for Weyl-interaction modified gravity, Int. J. Mod. Phys. D \textbf{26}, 1741001 (2017).
\bibitem{r24} S. Capozziello et al., Newtonian limit of $f(R)$ gravity, Phys. Rev. D \textbf{76}, 104019 (2007).
\bibitem{r25} M. Koussour and M. Bennai, tability analysis of anisotropic Bianchi type-I cosmological model in teleparallel gravity, Class. Quantum Gravity \textbf{39}, 105001 (2022).
\bibitem{r25a}S. S. Mishra, N. S. Kavya, P.K. Sahoo, V. Venkatesha, Constraining extended teleparallel gravity via cosmography: A model-independent approach, The Astrophysical Journal, \textbf{970}, 57 (2024).
\bibitem{r26} J. B. Jimenez et al., Coincident general relativity, Phys. Rev. D \textbf{98}, 044048 (2018).
\bibitem{r27}  J. B. Jimenez et al., Cosmology in 
f(Q) geometry, Phys. Rev. D \textbf{101}, 103507 (2020).
\bibitem{r28} Y. Xu et al., $f(Q,T)$ gravity, Eur. Phys. J. C \textbf{79}, 8 (2019).
\bibitem{snehafqt}Sneha Pradhan, Sunil Kumar Maurya, Pradyumn Kumar Sahoo, Ghulam Mustafa, Geometrically deformed charged anisotropic models in f(Q,T) gravity, Fortschritte der Physik  \textbf{72}, 2400092 (2024)
\bibitem{r29}
J.~Beltr\'an Jim\'enez, L.~Heisenberg and T.~Koivisto,
``Coincident General Relativity,''
Phys. Rev. D \textbf{98}, 044048  (2018).
doi:10.1103/PhysRevD.98.044048.


\bibitem{r30} R. Lazkoz, F.S.N.  Lobo, M. Ortiz-Banos,  V. Salzano, Observational constraints of $f(Q)$ gravity,  Phys. Rev. D, \textbf{100}, 104027 (2019).
\bibitem{r30a} R. Solanki, Avik De, P.K. Sahoo, Complete dark energy scenario in $f(Q)$ gravity, Physics of the Dark Universe \textbf{36}, 100996 (2022).
\bibitem{r31} S. Mandal, P.K. Sahoo, J.R.L. Santos, Energy conditions in $f(Q)$ gravity,  Phys. Rev. D, \textbf{102}, 024057 (2020).
\bibitem{r32} S. Mandal, D. Wang, P.K. Sahoo, Cosmography in $f(Q)$ gravity,  Phys. Rev. D , \textbf{102}, 124029 (2020).
\bibitem{r33} T. Harko  et al., Coupling matter in modified $Q$ gravity, Phys. Rev. D, \textbf{98}, 084043 (2018).
\bibitem{r33a}G. N. Gadbail, Sanjay Mandal, P.K. Sahoo, Gaussian Process Approach for Model-Independent Reconstruction of f(Q) Gravity with Direct Hubble Measurements, The Astrophysical Journal, \textbf{972}, 174 (2024).
\bibitem{r34} B. J. Barros et al., Testing $f(Q)$ gravity with redshift space distortions, Phys. Dark Universe, \textbf{30}, 100616 (2020).
\bibitem{r36} Z. Hasan,  S. Mandal, P.K. Sahoo, Traversable wormhole geometries in $f(Q)$ gravity, Fortschritte Der Phys. \textbf{69,} 2100023 (2021).
\bibitem{r37} R. Solanki, S.Mandal, P.K. Sahoo,Cosmic acceleration with bulk viscosity in modified $f(Q)$ gravity, Phys. Dark Universe \textbf{32}, 100820 (2021).
\bibitem{r38} D. Zhao, Covariant formulation of $f(Q)$ theory, Eur. Phys. J. C. \textbf{82}(4), 303 (2022).
\bibitem{r39} K. Hu,  T. Katsuragawa, and T. Qiu., ADM formulation and Hamiltonian analysis of $f(Q)$ gravity, Physical Review D \textbf{106},  044025 (2022).
\bibitem{r40} Gaurav N. Gadbail, S. Mandal, and P. K. Sahoo, Reconstruction of $\Lambda$CDM Universe in $f(Q)$ Gravity, Physics Letters B \textbf{835}, 137509 (2022).
\bibitem{r41} A. De,  and T.H. Loo, On the viability of $f(Q)$ gravity models, Classical and Quantum Gravity \textbf{40}, 115007 (2023).
\bibitem{r41a} S. Mandal and P.K. Sahoo, Constraint on the equation of state parameter ($\omega$) in non-minimally coupled $f(Q)$ gravity,  Physics Letters B \textbf{823}, 136786 (2021).
\bibitem{r41b} S. Mandal, Abhishek Parida, Pradyumn Kumar Sahoo, Observational Constraints and Some Toy Models in $f(Q)$ Gravity with Bulk Viscous Fluid, Universe \textbf{8}, 240 (2022).
\bibitem{r41c} W. Khyllep, et al.,Cosmology in $f(Q)$
 gravity: A unified dynamical systems analysis of the background and perturbations, Physical Review D \textbf{107}, 044022 (2023).
\bibitem{r42}  M. Koussour et al.,A New Parametrization of Hubble Parameter in $f(Q)$ 
 Gravity, Fortschritte der Physik \textbf{71}, 2200172 (2023).
\bibitem{r43}  S. Mandal, et al.,A study of anisotropic spheres in $f(Q)$
 gravity with quintessence field, Physics of the Dark Universe \textbf{35}, 100934 (2022).
\bibitem{r44} A. Banerjee, et al., Wormhole geometries in $f(Q)$ gravity and the energy conditions, The European Physical Journal C \textbf{81}, 1 (2021).
\bibitem{r45} G. Mustafa, Zinnat Hassan, P.K. Sahoo, Traversable wormhole inspired by non-commutative geometries in $f(Q)$ gravity with conformal symmetry, Annals of Physics \textbf{437}, 168751 (2022).
\bibitem{chodos}A. Chodos, R.L. Jaffe, K. Johnson, C.B. Thorn, V.F. Weisskopf,New extended model of hadrons, Phys. Rev. D  \textbf{9}, 3471 (1974).
\bibitem{Cheng1998} K.S. Cheng, Z.G. Dai, T. Lu, Strange stars and related astrophysical phenomena, Int. J. Mod. Phys. D \textbf{7}, 139 (1998).
\bibitem{Witten1984}E. Witten, Cosmic separation of phases, Phys. Rev. D \textbf{30}, 272 (1984).

\bibitem{Yan:2012mk}
Y.~Yan, J.~Cao, X.~L.~Luo, W.~M.~Sun and H.~Zong,
Connecting neutron star observations to the high density equation of state of quasi-particle model,
Phys. Rev. D \textbf{86}, 114028 (2012)
\bibitem{Schertler:1997za}
K.~Schertler, C.~Greiner, P.~K.~Sahu and M.~H.~Thoma,
The Influence of medium effects on the gross structure of hybrid stars,
Nucl. Phys. A \textbf{637}, 451-465 (1998)
\bibitem{Schertler:2000xq}
K.~Schertler, C.~Greiner, J.~Schaffner-Bielich and M.~H.~Thoma,
Quark phases in neutron stars and a 'third family' of compact stars as a signature for phase transitions,
Nucl. Phys. A \textbf{677}, 463-490 (2000)

\bibitem{Krori1975} K D Krori and J Barua,A singularity-free solution for a charged fluid sphere in general relativity, J. Phys. A: Math. Gen. \textbf{8} 508 ( 1975)

\bibitem{Ozel:2008kb}
F.~Ozel, T.~Guver and D.~Psaltis,
The Mass and Radius of the Neutron Star in EXO 1745-248,
Astrophys. J. \textbf{693}, 1775-1779 (2009)

\bibitem{Rawls:2011jw}
M.~L.~Rawls, J.~A.~Orosz, J.~E.~McClintock, M.~A.~P.~Torres, C.~D.~Bailyn and M.~M.~Buxton,
Refined Neutron-Star Mass Determinations for Six Eclipsing X-Ray Pulsar Binaries,
Astrophys. J. \textbf{730}, 25 (2011)

\bibitem{Herrera:1992lwz}
L.~Herrera,
Cracking of self-gravitating compact objects,
Phys. Lett. A \textbf{165}, 206-210 (1992)
\bibitem{Abreu:2007ew}
H.~Abreu, H.~Hernandez and L.~A.~Nunez,
Sound Speeds, Cracking and Stability of Self-Gravitating Anisotropic Compact Objects,
Class. Quant. Grav. \textbf{24}, 4631-4646 (2007)
\bibitem{Andreasson:2008xw}
H.~Andreasson,
Sharp bounds on the critical stability radius for relativistic charged spheres,
Commun. Math. Phys. \textbf{288}, 715 (2009)

\bibitem{Carvalho:2022kxq}
G.~A.~Carvalho, R.~V.~Lobato, P.~H.~R.~S.~Moraes, D.~Deb and M.~Malheiro,
Quark stars with 2.6 $M_\odot $ in a non-minimal geometry-matter coupling theory of gravity,
Eur. Phys. J. C \textbf{82}, no.12, 1096 (2022)
\bibitem{buch} H.A. Buchdahl, General Relativistic Fluid Spheres, Phys. Rev. D \textbf{116}, 1027 (1959)
\bibitem{48} N. Straumann, General Relativity and Relativistic Astrophysics (Springer, Berlin, 1984)
\bibitem{49} C.G. Bohmer, T. Harko,Bounds on the basic physical parameters for anisotropic compact general relativistic objects, Class. Quantum Gravity \textbf{23}, 6479 (2006)


\bibitem{Bhar:2023bcd}
P.~Bhar, K.~N.~Singh, S.~K.~Maurya and M.~Govender,
``A four parameters quark star in quadratic f(Q)\ensuremath{-}action,''
Phys. Dark Univ. \textbf{43} (2024), 101391
doi:10.1016/j.dark.2023.101391

\bibitem{Bhar:2024vxk}
P.~Bhar, A.~Malik and A.~Almas,
``Impact of f(Q) gravity on anisotropic compact star model and stability analysis,''
Chin. J. Phys. \textbf{88} (2024), 839-856
doi:10.1016/j.cjph.2024.02.016

\bibitem{Bhar:2023zwi}
P.~Bhar,
``Physical properties of a quintessence anisotropic stellar model in f(Q) gravity and the mass\textendash{}radius relation,''
Eur. Phys. J. C \textbf{83} (2023) no.8, 737
doi:10.1140/epjc/s10052-023-11865-5

\bibitem{Bhar:2023xku}
P.~Bhar and J.~M.~Z.~Pretel,
``Dark energy stars and quark stars within the context of f(Q) gravity,''
Phys. Dark Univ. \textbf{42} (2023), 101322
doi:10.1016/j.dark.2023.101322

\bibitem{Kaur:2024biz}
S.~Kaur, S.~K.~Maurya, S.~Shukla and B.~Dayanandan,
``Charged anisotropic fluid sphere in f(Q) gravity satisfying Vaidya - Tikekar metric,''
New Astron. \textbf{110} (2024), 102230
doi:10.1016/j.newast.2024.102230

\bibitem{Maurya:2024rlm}
S.~K.~Maurya, M.~K.~Jasim, A.~Errehymy, K.~S.~Nisar, M.~Mahmoud and R.~Nag,
``Singularity-Free Charged Compact Star Model Under F(Q)-Gravity Regime,''
Fortsch. Phys. \textbf{72} (2024) no.4, 2300229
doi:10.1002/prop.202300229

\bibitem{Maurya:2024dwn}
S.~K.~Maurya, A.~Aziz, K.~N.~Singh, A.~Das, K.~Myrzakulov and S.~Ray,
``Effect of decoupling parameters on maximum allowable mass of anisotropic stellar structure constructed by mass constraint approach in f(Q)-gravity,''
Eur. Phys. J. C \textbf{84} (2024) no.3, 296
doi:10.1140/epjc/s10052-024-12626-8

\bibitem{Deb:2018sgt}
D.~Deb, S.~V.~Ketov, S.~K.~Maurya, M.~Khlopov, P.~H.~R.~S.~Moraes and S.~Ray,
``Exploring physical features of anisotropic strange stars beyond standard maximum mass limit in $f\left(R,\mathcal {T}\right)$ gravity,''
Mon. Not. Roy. Astron. Soc. \textbf{485} (2019) no.4, 5652-5665
doi:10.1093/mnras/stz708
[arXiv:1810.07678 [gr-qc]].

\bibitem{Rahaman:2010mr}
F.~Rahaman, S.~Ray, A.~K.~Jafry and K.~Chakraborty,
``Singularity-free solutions for anisotropic charged fluids with Chaplygin equation of state,''
Phys. Rev. D \textbf{82} (2010), 104055
doi:10.1103/PhysRevD.82.104055
[arXiv:1007.1889 [physics.gen-ph]].

\bibitem{Kalam:2012sh}
M.~Kalam, F.~Rahaman, S.~Ray, S.~M.~Hossein, I.~Karar and J.~Naskar,
``Anisotropic strange star with de Sitter spacetime,''
Eur. Phys. J. C \textbf{72} (2012), 2248
doi:10.1140/epjc/s10052-012-2248-y
[arXiv:1201.5234 [gr-qc]].

\bibitem{Rahaman:2011hd}
F.~Rahaman, R.~Maulick, A.~K.~Yadav, S.~Ray and R.~Sharma,
``Singularity-free dark energy star,''
Gen. Rel. Grav. \textbf{44} (2012), 107-124
doi:10.1007/s10714-011-1262-y
[arXiv:1102.1382 [gr-qc]].

\bibitem{e1} O'Brien, S. and Synge, J.L. (1952) Jump Conditions at Discontinuities in General Relativity. Communications of the Dublin Institute for Advanced Studies, No. 9, 1-20.
\bibitem{e2} Robson, E. H. "Junction conditions in general relativity theory." Annales de l'institut Henri Poincaré. Section A, Physique Théorique. Vol. 16. No. 1. 1972.

\bibitem{Farhi:1984qu}
E.~Farhi and R.~L.~Jaffe,
Phys. Rev. D \textbf{30} (1984), 2379
doi:10.1103/PhysRevD.30.2379

\bibitem{hr1} Abbas, G., and H. Nazar. "Hybrid star model with quark matter and baryonic matter in minimally coupled f (R) gravity." Annals of Physics 424 (2021): 168336.
\bibitem{hr01} Nazar, H., and G. Abbas. "Model of charged anisotropic strange stars in minimally coupled f (R) gravity." Advances in Astronomy 2021, no. 1 (2021): 6698208.
\bibitem{hr02} Frolov, Andrei V. "Singularity problem with f(R) models for dark energy." Physical review letters 101, no. 6 (2008): 061103.
\bibitem{hr03} Reverberi, Lorenzo. "Curvature singularities from gravitational contraction in f(R) gravity." Physical Review D,  87, no. 8 (2013): 084005.
\bibitem{hr2} Saha, Pameli, and Ujjal Debnath. "Anisotropic quintessence strange stars in f (t) gravity with modified chaplygin gas." Advances in High Energy Physics 2018, no. 1 (2018): 3901790.
\bibitem{hr3} Camera, Stefano, Vincenzo F. Cardone, and Ninfa Radicella. "Detectability of torsion gravity via galaxy clustering and cosmic shear measurements." Physical Review D 89, no. 8 (2014): 083520.
\end{thebibliography}
\end{document}